\documentclass[a4paper,11pt,twocolumn,accepted=2021-03-19]{quantumarticle}
\pdfoutput=1
\usepackage{bm}
\usepackage{amsmath}
\usepackage{amssymb}
\usepackage{graphicx}
\usepackage{IEEEtrantools}

\usepackage{epsfig}
\usepackage{epstopdf}

\begin{document}

\newtheorem{theorem}{Theorem}
\newtheorem{corrolary}{Corollary}

\def\pr{\prime}
\def\be{\begin{equation}}
\def\en#1{\label{#1}\end{equation}}
\def\dag{\dagger}
\def\bar#1{\overline #1}
\def\U{\mathcal{U}}
\newcommand{\per}{\mathrm{per}}
\newcommand{\rd}{\mathrm{d}}
\newcommand{\vare}{\varepsilon }

 \newcommand{\m}{\mathbf{m}}
\newcommand{\n}{\mathbf{n}}
\newcommand{\s}{\mathbf{s}}
\newcommand{\bk}{\mathbf{k}}
\newcommand{\bl}{\mathbf{l}}
\newcommand{\br}{\mathbf{r}}

\newcommand{\Lim}[1]{\raisebox{0.5ex}{\scalebox{0.8}{$\displaystyle \lim_{#1}\;$}}}

\title{Distinguishing noisy boson sampling from classical simulations  }
 
\author{Valery  Shchesnovich }
\email{valery@ufabc.edu.br}

\affiliation{Centro de Ci\^encias Naturais e Humanas, Universidade Federal do
ABC, Santo Andr\'e,  SP, 09210-170 Brazil }

\begin{abstract} 
Giving   a convincing  experimental evidence of the  quantum supremacy  over classical simulations  is a challenging  goal. Noise is considered to be the main problem in such a demonstration, hence it is   urgent to understand  the effect of   noise. Recently found classical algorithms can  efficiently approximate,  to any small error, the output of  boson sampling  with   finite-amplitude  noise.  In this work it is shown analytically and confirmed by numerical simulations  that  one can  efficiently   distinguish   the output distribution of   such a  noisy   boson sampling    from   the approximations   accounting for  low-order quantum multiboson interferences,  what includes  the mentioned classical algorithms.  The number of samples  required to  tell apart the  quantum and classical output distributions   is strongly affected by the previously unexplored parameter: density of bosons,  i.e., the   ratio of   total number of interfering bosons  to  number of input ports of   interferometer.    Such critical dependence  is strikingly reminiscent  of the  quantum-to-classical transition in systems of  identical particles, which sets in    when the system size scales up while density of particles vanishes.      
\end{abstract}
\maketitle

\section{Introduction}
Quantum mechanics  promises  computational advantage   over  digital computers \cite{F,Sh}.   Current technology  is on the brink of building    quantum devices with the promised advantage   in  some specific  computational  tasks,   called the quantum supremacy \cite{P},  for which goal    several quantum systems are considered  \cite{AA,QSProp,QSArch,QSChar,QSColdAt}  and a dramatic breakthrough   was recently reported  \cite{GoogleS}.   Can  noise,   always present in an experimental setup,    compromise    the quantum supremacy  by allowing for an  efficient   classical simulation \cite{HQCF}?   In this work we consider how a noisy  boson sampling system can be distinguished from efficient classical approximations. 

In boson sampling  proposal of    Aaronson \& Arkhipov \cite{AA}    the  specific classically hard computational task is sampling   from  many-body   quantum interference of    $N$ indistinguishable bosons on a  unitary linear   $M$-dimensional interferometer. At least in the  so-called no-collision regime,   when  the output ports receive at most a singe boson (i.e., for $M\gg N^2$ \cite{Bbirthday}),   complexity-theoretic arguments have been found in Ref. \cite{AA} for the quantum advantage over classical simulations.    In general,   the output probabilities depend on the full many-body  quantum  interference of $N$ bosons given by a sum  of  $N!$ quantum amplitudes, i.e., by  the matrix permanents \cite{C,Scheel}, which are  hard to compute  \cite{Valiant,JSV,A1,Ryser}.

The boson sampling proposal had initiated efforts for experimental demonstration. Single photons  \cite{E1,E2,E3,E4,EVCC} as well as   Gaussian states  \cite{GBS,E5,GSNEW} in optical interferometers,  and   the  temporal-mode encoding     \cite{TbinBS,TbinBSExp} were proposed and tested for experimental implementation with quantum optics. Experimental quantum  optical platform  has seen significant advances  \cite{pure1phBS,HEBS,LossBS,12phBS}  culminating recently in an experimental implementation  with    20 photons  on a  60-mode interferometer   \cite{20ph60mod}.       Alternative  platforms  include   ion traps \cite{BSions}, superconducting qubits \cite{BSsuperc,Goldstein},   neutral atoms in optical lattices \cite{BSoptlatt}  and  dynamic   Casimir effect \cite{BScas}.   Initial estimate on the threshold size for demonstration  of quantum supremacy  with boson sampling   was  $N\approx 30$ bosons \cite{AA}. However,  recent   classical simulations  \cite{QSBS,Cliffords} pushed the     threshold  to  $N\approx  50$ bosons.   Inevitable experimental  noise  \cite{Goldstein,RR,KK,LP,VS14,Arkh,Brod,Latm}  additionally opens possibilities for  efficient classical approximation algorithms  \cite{K1,R1,OB,PRS,RSP,VS2019,NonUnifLoss}.  Most importantly,  if noise amplitudes do not vanish  when  the system  size scales up,   the recent algorithms of Refs. \cite{R1,RSP,VS2019}  approximate  efficiently the  output distribution of noisy boson sampling by low-order multi-boson interference. 

Unavoidable noise  has been  accounted for in the boson sampling  proposal  \cite{AA} by  allowing  for an approximation error.  Taking into account that the  complexity-theoretic arguments for the quantum supremacy are  asymptotic, whereas in an experiment one has a finite quantum device, one may wonder  how small experimental error should be?  In  the spirit of Ref.   \cite{F}, the experimental error problem can be formulated as follows: Can a classical algorithm efficiently  sample  from  the output distribution of a quantum device   in such a way that it would be   impossible to tell  from the sampling data whether we have the classical simulation  or  the quantum device?  At the  other  extreme,  the two output distributions  can be efficiently distinguished.  Since the promised  computational advantage  is due to  the opening gap  between  exponential and  polynomial  computations in the size of a quantum system  \cite{P},   it is reasonable  to consider a method of distinguishing a quantum device from a classical simulation as efficient if it  requires only  a  polynomial  in the  quantum system size number of samples.     
 
As a partial answer to the above problem, in the present work  it is shown that one can   efficiently distinguish the output distribution of   noisy  boson sampling   from output distributions of  a wide range of classical algorithms, such as   simulation with classical particles and the   recent algorithms of Refs. \cite{R1,RSP,VS2019}.    We give   an analytical  expression for the  {lower bound} on the total variation distance between the output distribution of  noisy boson sampling and  that of  such a   classical simulation and point  that   the probability of no particle counts in a subset of output ports can be used for  distinguishing the quantum and classical distributions.   The number of samples  necessary for distinguishing the quantum and classical distributions  critically  depends on  the  density of bosons, defined as the ratio of  the total number of interfering  bosons  to  the number of input ports of interferometer, and not on   the number of bosons or the number of ports of interferometer themselves. Our  analytical   results  are valid asymptotically with  the number of interfering bosons,   convergence to the asymptotic result  is  studied by using   numerical simulations.   

 Previously the output distribution of boson sampling   was shown    \cite{BSNotUn} to be far in  the  total variation distance from   the uniform distribution   (argued to be an efficient approximation \cite{BSUn}), where a set of open problems was given.  The present work  partially resolves   open  problems (2) and (4)-(6) of  Ref. \cite{BSNotUn} by considering  a  wide class of possible classical   approximations to   a noisy realization of  boson sampling with  arbitrary  scaling of the  interferometer size in the  total number of interfering  bosons, beyond the no-collision regime.   In  Ref.  \cite{BB} it was shown how one  can efficiently  distinguish the output distribution of  boson sampling from the  simulation by classical particles.    Statistical or pattern recognition techniques were  also  applied to assessment of boson sampling   and distinguishing it from classical simulations \cite{StatBench,ExpStatSign,WDBubles,PatternRec}. However,  no analysis of the impact of realistic experimental noise was attempted previously,  which is the  main  goal of the present work.
  
 The text is organized as follows. In section \ref{sec2}  the  boson sampling realization with linear optics and  the classical approximations by low-order multi-boson interference are described and  our  results on distinguishing  their respective output probability distributions are presented.      In section \ref{sec3} our  principal findings are summarized and discussed.    For better readability,  the  derivations  and mathematical details  are relegated to Appendices \ref{appA}-\ref{appF}.

\section{Noisy boson sampling vs classical approximations}
\label{sec2}

The boson sampling proposal of Ref. \cite{AA}   considers    the quantum   interference of   $N$ identical single  bosons on a  unitary $M$-port in the no-collision regime $M\gg N^2$, here we consider it  for arbitrary $M\ge N$.   It turns out that  distinguishing noisy   boson sampling,  as $N$ scales up,    crucially depends  on  density  of bosons  
  $\rho =N/M$,   more precisely, on the scaling $\rho =\rho(N)$ (or, equivalently, $M = M(N)$).  This fact allows us to consider all models with a given scaling  $\rho = \rho(N)$  as   a   class of boson sampling, where  $N$  serves as  the  size parameter  in the  class.   For example, for at least $\rho \sim 1/N$  we have the so-called no-collision regime, the main  focus of Ref.~\cite{AA}.       
   
  Let us  now  mention known estimates on  the  number of classical computations  required to simulate the  boson  sampling.   For the  ideal (noiseless) boson sampling, the  fastest to date algorithms of  Refs. \cite{QSBS,Cliffords} can sample  from  its   output distribution    in $O(N2^N)$ computations.  The number of  computations depends on the density of bosons: for non-vanishing density of bosons  the output probabilities  can be  given by the matrix permanents  of  rank-deficient matrices with  reduced computational complexity \cite{Barv,CompHard}.  The following simple rule  can be stated  \cite{MyEst}:   to sample from  the   boson sampling in arbitrary regime of density of bosons $\rho$    by the algorithm of  \cite{Cliffords}  requires at least as many   classical simulations  as for the  no-collision boson sampling   with $\mathcal{N}_\rho = N/(1+\rho)$ bosons.  The size parameter  $\mathcal{N}_\rho$ gives also the expected number of output ports occupied by bosons as a function of the boson density.  
  
Less is known about the  scaling of classical computations for exact   sampling from  \textit{noisy} boson sampling.  There are several strong   sources of  noise:  noisy interferometer   \cite{LP,Arkh}, partial  distinguishability of bosons \cite{VS14}, and boson losses \cite{Brod,Latm}.     It was shown that a finite number of lost bosons (or dark counts of detectors, or both)  does not compromise the  computational hardness of boson sampling  in the no-collision regime.    There are also  results for  the {approximate} sampling from noisy boson sampling, where the number of classical  computations depends additionally on the scaling of the amplitude of noise in the system size   \cite{KK,R1,RSP,VS2019}.    Efficient classical approximation  algorithm of  boson sampling  in  the no-collision regime with partially distinguishable bosons  was found  in Ref. \cite{R1}, then extended to losses  \cite{RSP} and noise in interferometer \cite{VS2019}.  For intermediate-size boson sampling devices    the number of computations of the classical algorithm can be  further  optimized  \cite{MRRT}.  There is equivalence of different imperfections in their effect on classical hardness of boson sampling,   e.g., noise in experimental platforms \cite{EVCC, Goldstein}  has  a similar effect   to  that of  the partial distinguishability of bosons  \cite{VS2019} (below we will use the term noise  for all imperfections).  
  
  The efficient classical algorithm of Refs.  \cite{R1,RSP,VS2019}  can approximate the output distribution  of $N$-boson sampling with a finite-amplitude noise by  a smaller  one, with  $K$ interfering bosons and $N-K$ classical particles, it becomes   efficient  (i.e., polynomial in $N,M$)  for bounded $K$, since  the  classical  computations scale  exponentially  only in  $K$ \cite{R1}.     Below we   investigate if  a noisy boson sampling with finite-amplitude noise   can  be efficiently distinguished from such a classical approximation, to answer the experimentally relevant problem posed in the previous section. 
    
\subsection*{Probability distribution at  output of  noisy boson sampling}

\begin{figure}[htb]
  \begin{center}
   \includegraphics[width=.5\textwidth]{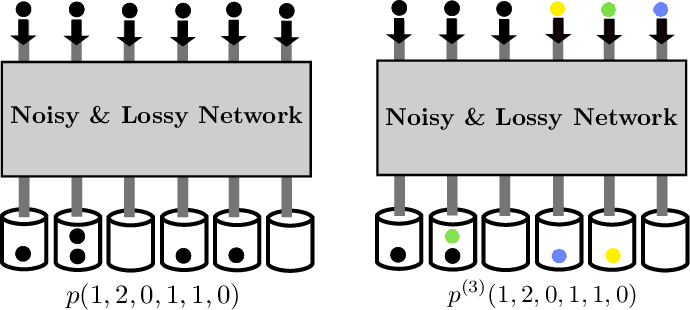}\\
           \caption{\textbf{ Noisy boson sampling vs classical approximation.}  Noisy $N$-boson quantum interference on a $M$-port    (left). An adversary   (right)  tries to approximate   the output distribution  of $N$-boson quantum interference,   $p(m_1,m_2,\ldots, m_M)$,  by   the output distribution, $p^{(K)}(m_1,m_2,\ldots, m_M)$, from  a combination of $K$-boson interference   (here $K=3$) and $N-K$ distinguishable bosons. }   \label{Fig1}
            \end{center}
                              \end{figure}

 Let us now briefly describe an experimental  implementation of boson sampling. We will use the popular example of  linear optical setup  with single photons, see fig.~\ref{Fig1},  and  focus on two   sources of noise: losses of photons and their partial distinguishability. Later on, we  will include also the false  random (a.k.a. dark) counts of detectors, i.e., the counts which are not triggered by photons.  There is an  equivalence  of the losses compensated by dark counts  with noise in the interferometer \cite{VS2019}, thus our model takes  into account the  strongest sources of  noise in linear  optical  experimental setup. The  multi-photon component at input is neglected here. It can be a strong  source of noise with the photon   sources based on the parametric down conversion,  e.g., when  about  $N^2$ sources are used   to produce $N$ single photons in the   boson sampling from a Gaussian state \cite{GBS,E5,12phBS}. With additional optical multiplexing of the sources one can  significantly reduce  the multiphoton component   noise  \cite{S2020}. 

Unitary  linear optical interferometer  $U$  with  $M$ input  and  output ports  connects  the optical   mode $|k^{(in)}\rangle$  of  input port  $k$    to   the optical modes $|1^{(out)}\rangle, \ldots, |M^{(out)}\rangle$ of the output ports:  $|k^{(in)}\rangle = \sum_{l=1}^M U_{kl} |l^{(out)}\rangle$. Such an interferometer would take a Fock state of   photons in the input ports  into a superposition of the Fock states of the  photons   in the output ports  according to the corresponding input-output relations between the boson creation operators   in the input and output ports:
 \be
 \hat{a}^\dag_{k} = \sum_{l=1}^M {U}_{kl}  \hat{b}^\dag_{l},
 \en{Eq00}
 where $\hat{a}^\dag_{k}$  ($\hat{b}^\dag_{l}$) is the  boson creation operator in input (output) port $k$ ($l$). The (photon number-resolving)  photon detection projects the output   state of   photons onto one of the Fock states in the output ports.   A realistic optical setup, however,    suffers  from photon losses and not all the   photons are detected.  Ways to compensate for   losses, such as by the post selection on a given number of photons  \cite{LossBS}, are currently under investigation.  For boson sampling one can allow only a  constant number  of bosons  to be lost \cite{Brod}. On the other hand,  if we assume that photons are lost independently of each other,  there will be,  on average,   a proportional to $N$ number of lost photons. The classical algorithm of Ref. \cite{RSP} can  efficiently approximate the   output distribution with the proportional losses. The post selection strategy  therefore can  work only for a small-size boson sampling, since the probability to lose a constant number of photons vanishes exponentially with $N$. Thus photon losses are unavoidable source of noise in an optical realization of boson sampling.   The input-output relation in the  case of arbitrary lossy interferometer $\mathcal{U}$,   can be cast in the following form \cite{LossNet} 
  \be
 \hat{a}^\dag_{k} = \sum_{l=1}^M\mathcal{U}_{kl}  \hat{b}^\dag_{l} + \sum_{j=1}^M V_{kj} \hat{c}^\dag_{j}, 
 \en{Eq0}
   where  $\hat{c}^\dag_j$ is the  boson creation operator in   loss mode $j$ (e.g., photon absorption due to non-unitarity of an interferometer), and matrix $V$ is such that the total unitarity is observed  $\mathcal{U}\mathcal{U}^\dag +VV^\dag =I$    (see more details in appendix \ref{appB}).

Another strong source of noise is partial distinguishability of photons, affecting the output state of photons on a linear interferometer \cite{HOM}.  Partial distinguishability can be associated with   different  internal states of photons  (such as temporal profile of a photon in case of Ref. \cite{HOM}),   which are unaffected by  the input-output relation in Eq. \eqref{Eq0} and  not resolved in an experiment.   In this case,  the boson operators in the input-output relations of  Eq. \eqref{Eq0}  refer  to  the same  internal state of boson on the input as well as on the output of the interferometer.  Partial  distinguishability of $N$ bosons can be  accounted for by  introducing  a   function  on permutations $\sigma\in S_N$, acting on the internal states  of bosons  \cite{VS14,PartDist}. This function is defined as follows. For pure internal states of bosons, $\hat{\rho}_k=|\psi_k\rangle\langle\psi_k|$, $k=1,\ldots, N$, it reads  
\be
J(\sigma) = \prod_{k=1}^N \langle \psi_{\sigma(k)}|\psi_{k}\rangle.
\en{Jdef}
 For   mixed internal states  $J(\sigma)$ is a convex sum of the products as in Eq.~\eqref{Jdef}.   Note that   $J(\sigma)$   of Eq. \eqref{Jdef} factorizes  \mbox{$J(\sigma) =   J(\mu_1)J(\mu_2) \ldots J(\mu_q)$} according to the disjoint cycle   decomposition $\sigma = \mu_1\mu_2\ldots \mu_q$  of permutation $\sigma$ \cite{Stanley}, where each  cycle  \mbox{$\mu: k_{1}\to k_{2}\to  \ldots \to k_{{|\mu|}}\to k_{1}$} ($|\mu|$ denotes  the cycle length) contributes a   factor $J(\mu)$ given by a similar expression as  the right hand side of Eq. \eqref{Jdef} with $\sigma$ replaced by $\mu$. In general, each cycle-factor $J(\mu)$ accounts   for a specific $|\mu|$-boson interference process  \cite{Ninter} and the factorization of $J$ occurs    when bosons are uncorrelated, i.e., when  their  internal state is  factorized $\hat{\rho}_1\otimes\ldots\otimes \hat{\rho}_N$.   For instance,  in   the famous $2$-photon  interference experiment  \cite{HOM},  for the transposition of the  photons (i.e., $2$-cycle  $1\to 2\to 1$) we have       
 \mbox{$J = |\langle\psi_1|\psi_2\rangle|^2$}, with $|\psi_k\rangle$ being  the  temporal shape of  photon $k$.  This value  enters the probability of the coincidence count at the output of a unitary linear $2$-port: $p(1,1) = (1-J)/2$, i.e., the Mandel dip in Ref. \cite{HOM} (more on the partial distinguishability  can be found in Ref. \cite{PartDist}). Explicit  analytical results below  are obtained for the uniform partial distinguishability,  controlled  by a single  distinguishability  parameter $0\le \xi\le1$. This model applies when boson $k$  with probability $\xi$ is in   some   pure internal state $|\phi_0\rangle$ and with probability $1-\xi$ in an unique internal state $|\phi_k\rangle$, orthogonal to all other internal states. Either pure $\hat{\rho}^{(a)}_k $  or  mixed $\hat{\rho}^{(b)}_k$ state of boson $k$  may correspond to such a  case:
\begin{IEEEeqnarray}{llC}
  \hat{\rho}^{(a)}_k = |\psi_k\rangle \langle\psi_k|, \quad |\psi_k\rangle = \sqrt{\xi}|\phi_0\rangle + \sqrt{1-\xi}|\phi_k\rangle, &  \nonumber \\
   \hat{\rho}^{(b)}_k = \xi |\phi_0\rangle\langle\phi_0| + (1-\xi)  |\phi_k\rangle\langle\phi_k|, &  \nonumber\\
   \langle \phi_k|\phi_j\rangle  = \delta_{kj}, \quad k,j\in\{0,1,\ldots, N\}. & 
 \label{IntSt}
 \end{IEEEeqnarray}
For  bosons  in internal  states $\hat{\rho}^{(a,b)}_1, \ldots, \hat{\rho}^{(a,b)}_N$  of  Eq. \eqref{IntSt},   each $\mu$-cycle for $|\mu|\ge 2$ contributes  the factor $\xi^{|\mu|}$ to the  distinguishability function in Eq.~\eqref{Jdef}.  The sum of all the cycle lengths  in a permutation $\sigma$ satisfies  $\sum|\mu| = N-c_1(\sigma)$,   where  $c_1(\sigma)$ is the number of fixed points ($1$-cycles),  hence  $J(\sigma) = \xi^{N-c_1(\sigma)}$.  If photon     detectors do not resolve  the   internal states of photons,  the    same distinguishability function corresponds to  pure-state  or mixed-state  model  in Eq. \eqref{IntSt}.     In an experiment, there is always noise in photon parameters,  such as random fluctuations of photon  arrival times,  thus the internal states of photons are always mixed.  Whatever is the  model of photon detection, the distinguishability of photons due to mixed states  will persist, since    even by completely  resolving the internal states of photons one  is not  able to affect the  distinguishability coming from the  fluctuations of photon parameters    \cite{VS2020}. Thus partial distinguishability of photons is a source of unavoidable noise in  boson sampling experiments with  photons.

We can now give the output distribution of our   noisy boson sampling model with  boson losses and  partially distinguishable  bosons. When $N$ bosons, with the distinguishability function $J(\sigma)$,  are sent to the input ports $k=1,\ldots, N$ of a lossy interferometer  $\mathcal{U}$, the probability to detect exactly $n$ of them, \mbox{$0\le n\le N$},  at the output ports    in a configuration $\m= (m_1,\ldots,m_M)$, $m_1+\ldots+m_M=n$, i.e., when $m_l$ bosons are detected at output port $l$,    reads (see appendix \ref{appB} for details)
\begin{widetext}
\begin{align}
\label{Eq1}
 p(\m) =\frac{1}{\m!}\sum_{\sigma\in S_N} J(\sigma)\sum_{\bk}\sum_{\tau\in S_n}
 \prod_{\alpha=1}^n \mathcal{U}^*_{\sigma\tau(k_\alpha),l_\alpha} \mathcal{U}_{\tau(k_\alpha),l_\alpha}   \prod_{\alpha=n+1}(I- \mathcal{U}\mathcal{U}^\dag)_{k_\alpha, \sigma(k_\alpha)},
\end{align}
\end{widetext}
where   the sum over $\bk=(k_1,\ldots,k_n)$ stands for summation over all  $n$-dimensional subsets of $\{1,\ldots, N\}$,   the multi-set $l_1,\ldots, l_n$ contains   the  output ports  with  bosons  (one for each detected boson)  and $\m! = m_1!\ldots m_M!$.

When only $n<  N$ bosons   are detected at the output,  there is still some  interference  between the $N-n$ lost bosons and the  $n$  detected ones,  reflected  in exchange of bosons between the output and the loss modes by permutation $\sigma$ in Eq. \eqref{Eq1}. Only for a diagonal loss  matrix  $I-\mathcal{U}\mathcal{U}^\dag$  no such boson exchange occurs (in this case $\sigma(k_\alpha) = k_\alpha$ for all \mbox{$n+1\le \alpha\le N$}) and the residual interference disappears. In the latter case, boson   losses can be  considered to occur, e.g.,  at the  input of the  interferometer  \cite{PRS} and there is such $D = \mathrm{diag}(\eta_1,\ldots,\eta_M)$, $0\le \eta_k\le 1$, that    $\mathcal{U} = \sqrt{D} U$ for some $UU^\dag= I$. Parameter  $\eta_k$ in this case is the probability that  boson sent to input port $k$ actually passes the interferometer  (i.e., $1-\eta_k$ is the probability that boson $k$ is lost).  Below, some of our explicit results  are given for  the uniform loss model with $\U = \sqrt{\eta} U$, $U^\dag U = I$, where $\eta$ is the uniform transmission of such an  interferometer.

\subsection*{ Approximations by $K$ interfering bosons  and $N-K$ classical particles}

In the no-collision regime, the  output distribution of noisy $N$-boson  sampling model    can be approximated, to any small error,   by   a similar noisy  model with  bosonic particles,  where there are only $K$ interfering bosons and the rest $N-K$  particles are  distinguishable bosons (i.e., classical particles)  \cite{R1,RSP,VS2019}. The  parameter $K$ is chosen according to the required approximation  error.  Below we adopt   such a classical approximation as the classical adversary, see fig.~\ref{Fig1}.

The simplest way to  introduce the above described  classical  approximation is  via the respective distinguishability function  $J^{(K)}(\sigma)$, which is  equal to $J(\sigma)$  for permutations  $\sigma$ with at least $N-K$ fixed points and   to  zero otherwise \cite{VS2019}: 
\be
J^{(K)}(\sigma) \equiv \left\{\begin{array}{cc}J(\sigma), & c_1(\sigma) \ge N-K,\\ 0, & c_1(\sigma)< N-K.  \end{array}   \right.
\en{Eq2}
With such a distinguishability function, for $K>1$,   $N-K$ randomly chosen    bosons   contribute to the output probability  in the same way as classical particles, whereas the rest $K$ bosons are allowed to interfere.    By setting   $K=1$  or   $K=0$  we get     $N$ classical particles (since a permutation   with   $N-1$ fixed points is the identity permutation). 

Our goal is to estimate the total variation distance between the output distributions of noisy   boson sampling (${p}$)  and  a classical    approximation (${p}^{(K)}$), where both distributions are given by   Eq. \eqref{Eq1} with the corresponding distinguishability functions  $J(\sigma)$   and $J^{(K)}(\sigma)$. The total variation distance    is  defined as follows 
\be
 \mathcal{D}(p,p^{(K)})= \frac12 \sum_{\m} |p(\m) - p^{(K)}(\m)|, 
\en{Dlower}
where the sum runs over all possible  configurations $\m=(m_1,\ldots,m_M)$ of  bosons  in the output ports.    The key observation   is that  for any subset $\Omega$ of  the  configurations  of bosons 
in the output ports we have 
\be
\mathcal{D}(p,p^{(K)})\ge \left|  {P}_\Omega-P^{(K)}_\Omega \right|, \quad P_\Omega \equiv \sum_{\m\in \Omega} p(\m).
\en{Dbound}
Observe that the  equality   is necessarily achieved for a certain subset 
$\Omega_*$  depending  on $\U$ and other parameters of the setup. 

Analytical analysis below is  carried out for the  difference  in  probability, $\Delta P_L= P_{\Omega_L} - P^{(K)}_{\Omega_L}$, of all the  output bosons to be detected in a  subset of   $M-L$   output ports, or, equivalently,  no particle counts  in the complementary   subset of the output ports $\Omega_L \equiv \{l_1,\ldots,l_L\}$. Let us assume  that such a subset is chosen once for a given setup (below we will return to  this point in more detail).   Since we   consider arbitrary (or randomly chosen multiports) we will use  $\Omega_L = \{1,\ldots, L\}$ in our analytical and numerical considerations below.  

At this stage, the dark counts of detectors  can be easily   accounted for.    Dark counts of a detector  follow Poisson distribution $p_d(n) = \frac{\nu^n}{n!}e^{-\nu}$, where we assume a uniform   rate $\nu$ for all detectors.  Hence, the dark counts contribute   the   factor $e^{-L\nu}$  to the  the  probability  of no  counts  in $L$ output ports.   Denoting $\Delta J = J - J^{(K)}$ we get from Eqs. \eqref{Eq1}-\eqref{Eq2}   the following expression  for the difference in probability of no counts in $L$ output ports  (see details in appendices \ref{appA} and \ref{appB})
\begin{IEEEeqnarray}{CCl}
\label{Eq3}
&  \Delta P_L  = e^{-L\nu}\sum_{\sigma\in S_N}   \Delta J(\sigma)\prod_{k=1}^N A_{k,\sigma(k)},  \\
&   A_{kj}\equiv \delta_{kj}  - \sum_{l\in \Omega_L} \mathcal{U}_{kl}\mathcal{U}^*_{jl}.\nonumber
\end{IEEEeqnarray}

In the limit of  large number of interfering bosons $N\gg 1$, the lower bound on the total variation distance between a noisy boson sampling and the classical approximation by $K$ interfering bosons,  Eq.~\eqref{Dbound},   depends on the   density  of bosons $\rho=N/M$,  and not on the number of interfering bosons $N$ and the size $M$ of interferometer themselves.   We will prove this in the case of interferometer with uniform transmission $\eta$, $\mathcal{U} = \sqrt{\eta }U$, $U^\dag U =I$, and  uniform distinguishability $\xi$,  i.e., with for the distinguishability function $J(\sigma) = \xi^{N-c_1(\sigma)}$.      For  $L=1$ the   difference in probability becomes 
\begin{align}
\label{DP1}
  \Delta P_1  &= e^{-\nu }  \sum_{n=K+1}^N (-\eta)^n  \sum_{\bk}\prod_{\alpha=1}^n |U_{k_\alpha,1}|^2  \nonumber\\
 & \times \sum_{m=K+1}^{n}\binom{n}{m}d_{m} \xi^{m}, \nonumber\\
 &d_{m} \equiv m! \sum_{s=0}^{m}\frac{(-1)^s}{s!},  
 \end{align}
 where  the summation over $\bk=(k_1,\ldots, k_n)$   runs over all $n$-dimensional subsets of $1,\ldots, N$  (the derivation can be found in appendix \ref{appC}). In Eq. \eqref{DP1}  the first two  sums  are  due to expansion of the loss matrix $A_{kj} = \delta_{kj} - \eta U_{k1}U^*_{j1}$ of Eq. \eqref{Eq3} in powers of $\eta$, while the third sum gives the  partial distinguishability of the   $n$-boson interference: there are  $\binom{n}{m}d_m$ amplitudes of $m$-boson interferences \cite{VS2020},  weighted by  $\xi^m$,  where  $d_m$ is  the total number of $m$-dimensional derangements (permutations with no fixed points) \cite{Stanley} and $\binom{n}{m}$ is the number of $m$ dimensional subsets of $n$ bosons.  Such interferences up to the order $m\le K$ are accounted for  by our  classical model, hence the summations in Eq. \eqref{DP1} start from $ K+1$.   

Eq. \eqref{DP1}   and the known average \cite{HaarR}  over the Haar-random interferometer $U$,   
\[
\langle |U_{k_1,1}|^2\ldots |U_{k_n,1}|^2\rangle =    \frac{1}{M^{(n)}},
\]
where $M^{(n)} = M(M+1)\ldots (M+n-1)$, allow us to  easily find the average value
\be 
\langle  \Delta P_1 \rangle  = e^{-\nu}\!\!  \sum_{n=K+1}^N \binom{N}{n}  \frac{(-\eta)^n}{M^{(n)}} \!\!\sum_{m=K+1}^{n}\binom{n}{m}d_{m} \xi^{m}.
  \en{<DP1>}
Similarly,  one can derive an expression   for the variance $\langle (\Delta P_1)^2 \rangle  - \langle \Delta P_1\rangle^2$ (see appendix \ref{appC}). 
  
 \begin{figure}[ht!]
      \includegraphics[width=.475\textwidth]{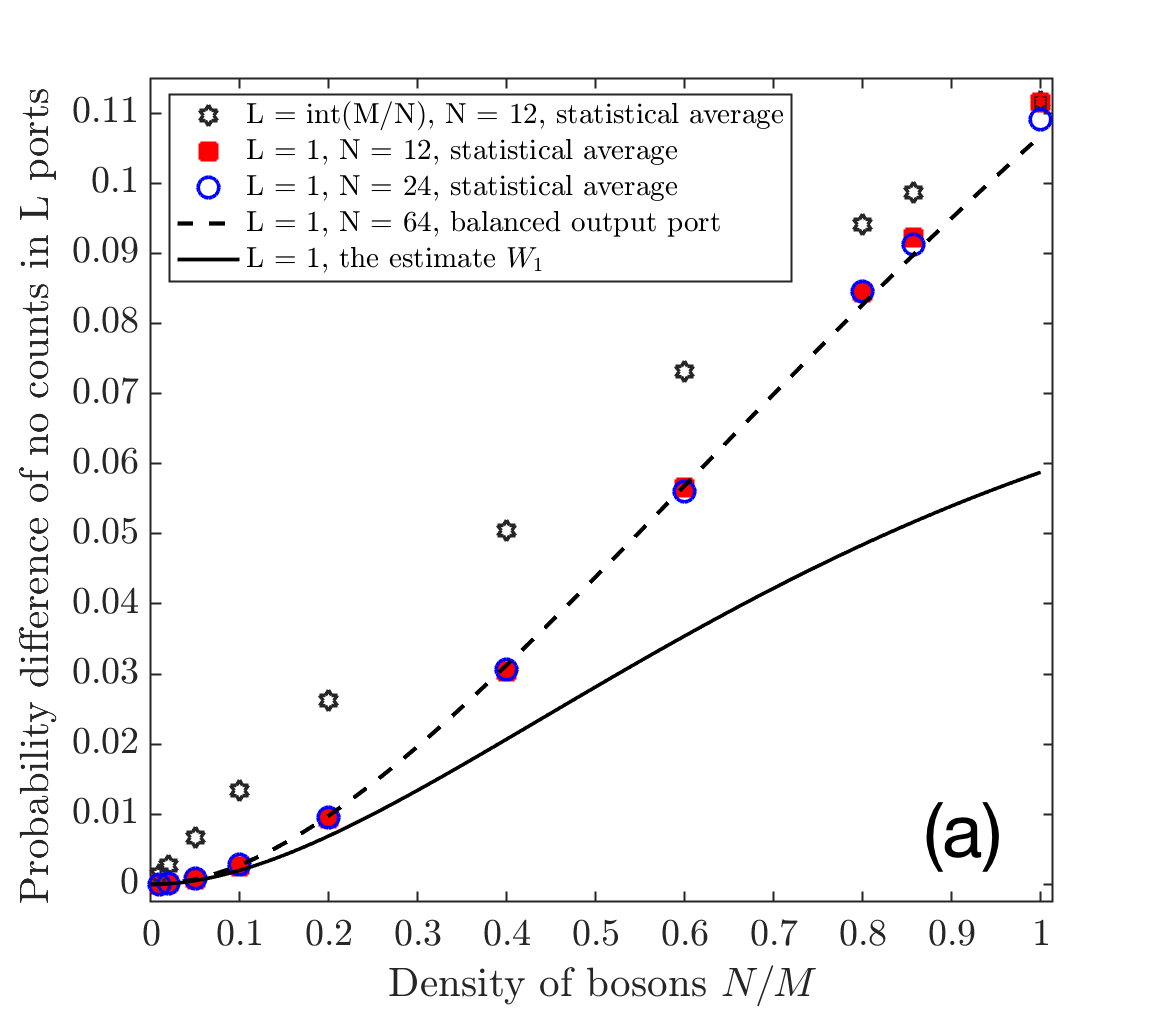} 
        \includegraphics[width=.475\textwidth]{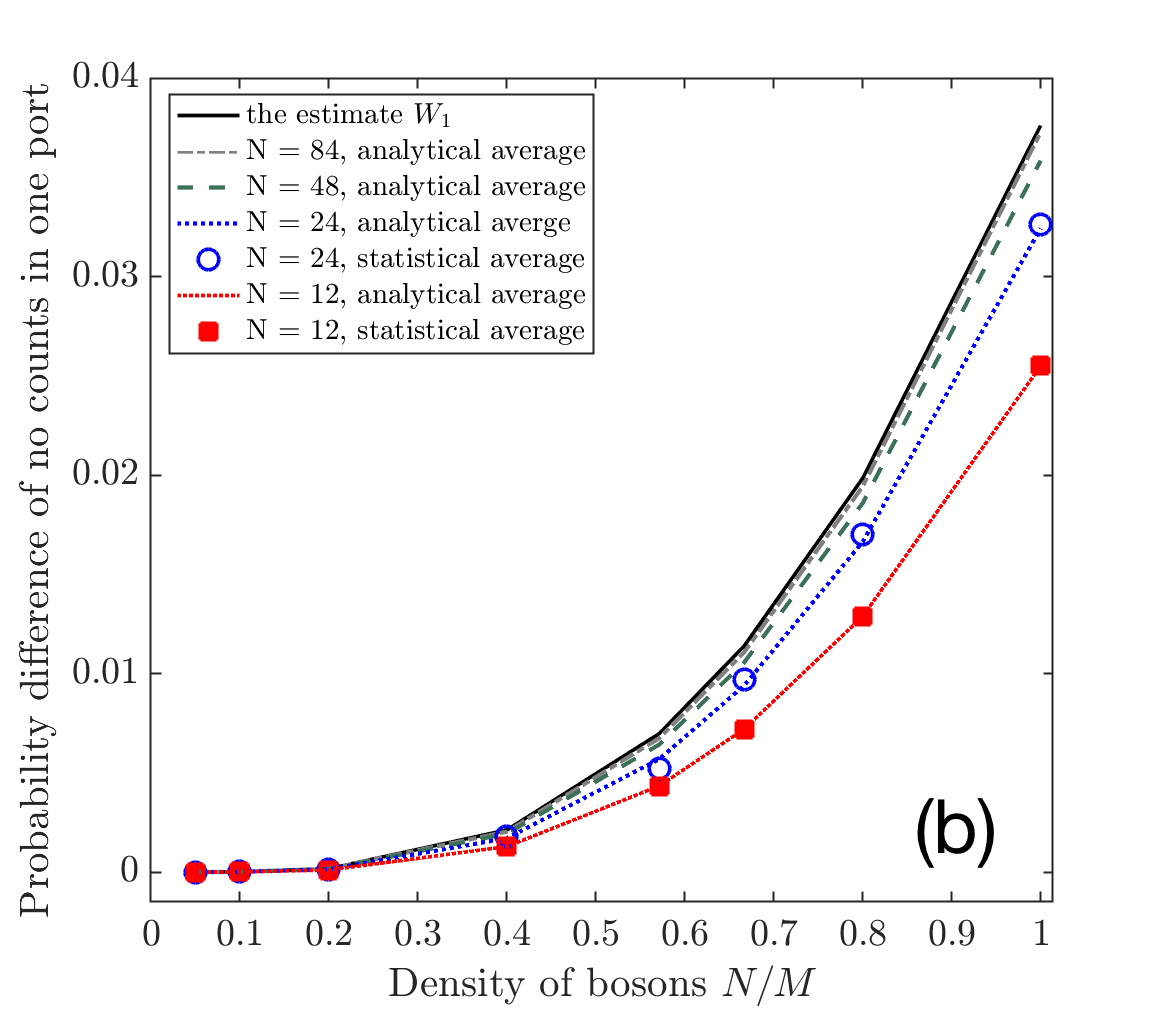}
          \caption{\textbf{The lower bound on the total variation distance.} Uniform  boson losses, with  the transmission    $\eta = 0.8$, serve  as  noise.  In panel (a)  the classical simulation  is by distinguishable bosons ($K=1$), whereas in panel (b)  by the third-order multi-boson  interferences ($K=3$).   In panel (a) and (b) the solid lines give   the  asymptotic analytical estimate $W_1$,   the squares ($N = 12$) and circles ($N=24$)  give the difference in probability of no counts in a single   output port, obtained by statistical averaging over random multiports. In panel (a)  the dashed line gives the difference in probability of no counts  in a  balanced output port and  the hexagons  give the statistical average of the   difference in probability of no counts  in $L  = \mathrm{int}(M/N)$ output ports.  In panel (b) the dotted lines compare the  analytically obtained  average   with the statistical average for the same $N$, the dashed line gives the analytical average for $N=48$,  and the dash-dotted  that  for $N=84$.}
          \label{fig2}
   \end{figure}
For   $K\ll \sqrt{N}$    the difference in probability   in Eq. \eqref{<DP1>}  becomes a function of     
$K$, the amplitudes of noise, $\xi,\eta,\nu$ and   the density of bosons $\rho=N/M$.   To extract the dependence on the setup parameters from  Eq.~\eqref{<DP1>},  we use   that  $d_{K+i} =  (K+i)!  \left(e^{-1} +R_{K+i}\right) $ with  $|R_m| < 1/(m+1)!$ (see   appendix \ref{appF})  and, assuming that $K$ is sufficiently large,   drop the remainder $R_{K+i}$.    By expanding  the resulting expression    in  powers of  $K^2/N$    and retaining only the leading order terms we obtain (see appendix \ref{appC}):
\begin{align}
\label{Eq4}
    &|\langle \Delta P_1\rangle | \approx   \frac{(\xi \eta\rho)^{K+1}}{1+\xi \eta \rho}e^{-1-\nu-\eta \rho}\equiv W_1,  \\
   &\frac{\langle (\Delta P_1)^2 \rangle  - \langle \Delta P_1\rangle^2}{W^2_1} \approx  \frac{(1-\rho)(K+1)^2}{N}. \nonumber 
\end{align}
Furthermore, for the   class of interferometers having   one balanced output port $|U_{k,l}| = \frac{1}{\sqrt{M}}$  one can show that the difference in probability of no counts   in the balanced output port  satisfies  
\be
|\Delta P_1|\ge W_1\left[1- \left|O\left( \frac{K^2}{N}+   \frac{1}{(K+1)!}\right)\right|\right],
\en{LowBound}
 where the  minus sign indicates a negative correction (see  details in appendix \ref{appC}). Such interferometers contain  a wide class: $U= \mathcal{F}(1\oplus  V)$ with the Fourier interferometer  $\mathcal{F}_{kl} =\frac{1}{\sqrt{M}}e^{2i\pi\frac{kl}{M}}$ and an arbitrary  $(M-1)$-dimensional unitary interferometer $V$.  

By considering more than one output port for the same setup  one can maximize   $\left|\langle  \Delta P_L\rangle \right|$   as a function of  $L$  (i.e., the lower bound on the total variation distance can be optimized). Indeed,   for noiseless  boson sampling,   $\xi=\eta=1$ and $\nu = 0$,    for   classical particles   ($K=1$) and   $L\ll M$  we have     $ \langle \Delta P_L \rangle  =  (1+\rho)^{-L} - e^{-L\rho}$    \cite{VS2017}, where the average is  over the Haar-random  unitary interferometer, maximized    when $L = \mathrm{int}(1/\rho)=\mathrm{int}(M/N)$.

The  analytical results of Eqs. \eqref{DP1}-\eqref{LowBound} were  verified  by numerical simulations of  Eq. \eqref{Eq3} for  the difference in probability of no counts in $L\ge 1$ output ports  (the numerical approach is described in  appendix \ref{appD}).  The results are plotted in fig. \ref{fig2}.   In the simulations the uniform boson  losses with the transmission $\eta = 0.8$,  $\mathcal{U}=\sqrt{\eta} U$, is the only source of noise  (i.e., bosons are completely indistinguishable, $\xi=1$,  and there no dark counts of detectors, $\nu=0$).  Averaging over  the   unitary matrix $U$ by selecting it  uniformly randomly  from the unitary group    we obtain what is called in fig.~\ref{fig2}  the ``statistical average".   Our  method  allows such averaging only   for small number of bosons $N\le 24$,  due to necessity of computing a large number of $N$-dimensional matrix permanents.  Therefore, for   $N\le 24$ and $L=1$ we  check against the statistical average  the  analytical expression of Eq. \eqref{<DP1>}, called  in fig.~\ref{fig2}  the ``analytical average", which is then used for larger numbers of bosons.

The numerical results confirm  the scale-invariance  predicted by Eq.~\eqref{Eq4}, when the applicability condition \mbox{$K\ll \sqrt{N}$} is satisfied:  in  fig.~\ref{fig2}(a) (where $K=1$) the   statistical average values   $|\langle \Delta P_1\rangle|$  with   $N=12$ and $N = 24$  interfering bosons are  almost indiscernible, whereas in   fig.~\ref{fig2}(b) ($K=3$)  the numerical results gradually  approach our  asymptotic  estimate $W_1$ as  the number of interfering bosons $N$  scales up from $N=12$ to $N=84$.   Similar convergence of  $|\langle \Delta P_1\rangle |$ to  $W_1$  as $N$ scales up was observed numerically  for  other values  $K> 3$.  The opening gap  between $W_1$ and $|\langle \Delta P_1\rangle |$ in fig.~\ref{fig2}(a)  results from  the approximation for the number of derangements $d_{K+i}\approx e^{-1}(K+i)!$ (see Eq. \eqref{DP1}) used  in  Eq. \eqref{Eq4},  not justified  for  $K=1$.

Finally, our results  agree with  the \textit{upper bound} on the total variation distance in the no-collision regime. A   noisy boson sampling setup  of arbitrary  large size \mbox{$N\gg 1$}   can be  approximated in the no-collision regime   by the    distribution $p^{(K)}$  \cite{R1,RSP,VS2019} to the    total variation distance     \mbox{$\mathcal{D}(p,p^{(K)})=  O\left((\xi\sqrt{\eta})^{K+1}\right)$}. Whereas the upper bound is independent of $N$ in this regime, our  asymptotic \textit{lower bound}  $W_1$ of Eq. \eqref{Eq4} vanishes  due to dependence on the density of bosons   $W_1\sim \rho^{K+1}$,  since  in the no-collision regime  $\rho \ll 1/N$.  No upper bound is known beyond the no-collision regime.

\subsection*{ The set of efficient distinguishers} 
 
The above results  not only show that one    \textit{can} efficiently distinguish the output distribution of  a   noisy boson sampling device  from that of a classical approximation,   such  as those used in  Refs. \cite{R1,RSP,VS2019}, but  provides  a   large set of such distinguishers.  The probability $P_L$ of detecting zero counts in $L$ output ports, used  in    Eq. \eqref{Eq3}, is such a  distinguisher. Note that there  are     $\binom{M}{L}$ of ways to select a subset $\Omega_L$ with $L$ ports from the total $M$ of them. There is enough  subsets   $\Omega_L$  (with different  $L$ and $\Omega_L$) with independent probabilities $P_L = P_L(\Omega_L)$~\footnote{For example, one  sufficient number of subsets  is obtained by   fixing  $L<M-N$ and varying    $\Omega_L$. Note, however, that  one cannot sample from the  output distribution $p(\m)$  using the sufficient set of $P_{\Omega_L}$.} for  inversion of   the relation  $P_L = \sum_{\Omega_L} p(\m)$,  where $p(\m)$ is the output distribution of boson sampling, thus  there is a strong   correlation between the set of  probabilities $P_L(\Omega_L)$ and the output distribution. 

For   uniform distinguishability $J(\sigma) = \xi^{N-c_1(\sigma)}$, uniform  detector dark counts rate $\nu$,  and arbitrary lossy interferometer $\mathcal{U}$  we get $P_L$ as a single matrix permanent   of $N$-dimensional positive-semidefinite Hermitian matrix 
\begin{align}
\label{Eq5}
 P_L = e^{-L\nu} \mathrm{\per}A(\xi), \quad 
   A_{kj}(\xi) \equiv \left\{ \begin{array}{cc} A_{kk}, j=k,\\ \xi A_{kj}, j\ne k,\end{array} \right. 
\end{align}
where $A$ is from Eq. \eqref{Eq3}.  Such permanents can be efficiently approximated  by a quantum-optics inspired algorithm \cite{QuantOptPerHerm,PermThermStates}.

There is also an analytical expression for the  average  probability of no counts  $P_1$  in a single output port of our noisy boson sampling   with uniform distinguishability, uniform  losses and dark counts (details in appendix \ref{appE}),
\be
\langle P_1 \rangle = e^{-\nu} \sum_{n=0}^N    \frac{(N)_n}{M^{(n)}} (-\eta)^n\sum_{s=0}^n \frac{\xi^{n-s}(1-\xi)^s}{s!},
\en{P_1}
where $(N)_n = N(N-1)\ldots (N-n+1)$. Moreover, for $N\gg 1$ and  $\rho\eta\ll 1$ one can approximate  (see appendix \ref{appE}): 
\begin{align}
\label{EqP_1}
&  \langle P_1\rangle  \approx  \frac{\exp\left(-\nu-\rho\eta[1-\xi]\right)}{1+\xi\rho\eta}, \\
& \frac{\langle (P_1\! -\! \langle P_1\rangle )^2\rangle}{\langle P_1\rangle^2}  \approx  \frac{(\eta\rho)^2(1\!-\!\rho)}{N} \left(1-\xi + \frac{\xi}{1+\rho\eta\xi} \right)^2.\nonumber
\end{align}

In an experiment, there is no need to   actually compute   $P_L$: one has only to  choose $L$ output ports    and estimate  $P_L$   from   the number of output data  with zero counts in them.    Let us  analyze  an example.   According to the standard estimation theory (e.g., Ref. \cite{EstTheory}),    the probability $P_1$ can be  estimated, with  $\alpha$-confidence level,  as  \mbox{$P_1^{(E)} = (T_S+z^2_\alpha/2)/(T+z^2_\alpha)$,} where  $T$ is the total number of samples, $T_S$ is the number of successes (samples  with zero counts in  the selected output  port) and $z_\alpha$ is the quantile of the standard normal distribution (e.g., for $\alpha = 0.05$,  $95\%$-confidence, $z^2_\alpha\approx 4$).  For   
\[
T =  P_1(1-P_1)\left(\frac{1-\alpha/2}{\epsilon W_1}\right)^2
\]
 samples    the relative  confidence interval for $P_1$ becomes  $\frac{|P_1- P_1^{(E)}|}{W_1} = \epsilon$.    Setting    $\epsilon   =\alpha$,  with the same confidence level as that of  the estimate on $P_1$, by experimentally estimating $P_1$ one can tell the output  of   the quantum device  from that of the    classical simulation.   For $K = O(1)$  and bounded amplitudes of noise,  the  number of samples sufficient  for telling apart the two outputs  depends on  $N$  only through the density of bosons  $\rho=N/M$ (see Eq. \eqref{Eq4}).  Hence, for  any scale $M\sim N^{\kappa}$, $\kappa \ge 1$, it  is polynomial in $N$, which by our definition means  efficiency.  Interestingly, the required number of samples   remains  bounded as $N$ scales up  if  density of bosons remains finite, i.e., for $M\sim N$.

 \subsection*{  Approximations accounting for \textit{all} $K$-boson interferences}
 
 One important observation on the choice of our classical model, Eq. \eqref{Eq2},  is in order. The  highest order of quantum multiboson interference accounted by   the  classical approximation in Eq. \eqref{Eq2} is obviously $K$. However, by allowing only $K$ bosons to interfere,  such a  model does not account for the multi-boson interference where groups of   up to $K$ bosons interfere between themselves. Such an interference corresponds to a subset of  permutations in the permutation group, which decompose into the disjoint  cycles of  length  up to  $K$.   Thus, if  one wants to take into account    \textit{all} the multiboson interferences  up to the $K$th order, one has to use a different   model \cite{Ninter}, say $J^{(K)}_+$, obtained by setting  to  zero the distinguishability function $J(\sigma)$ on all  permutations $\sigma$  having cycles of length $n\ge K+1$ in the disjoint   cycle decomposition: 
\be
J_+^{(K)}(\sigma) \equiv   \left\{\begin{array}{cc}J(\sigma), &  c_{K+i}(\sigma)=0, \forall\;  i\ge 1,\\ 0, & \exists\; c_{K+i}(\sigma)>0,  \end{array}   \right.
\en{JKnew} 
where $c_{n}(\sigma)$ is the number of cycles of length $n$ in the cycle decomposition.  

However,       switching to the model of Eq. \eqref{JKnew}, brings  insignificant changes to the results, e.g., changes only  the     $K$-independent factor   in $W_1$  of Eq. \eqref{Eq4} (see details in appendix \ref{appF}).   Our model   with $J^{(K)}$ of  Eq. \eqref{Eq2}  simplifies the calculations   and allows for a simple numerical  algorithm  (see appendix \ref{appD}), whereas producing  essentially the same    lower bound  on the total variation distance as the model  in  Eq. \eqref{JKnew}.  This  fact leads to two important conclusions:   (i) our approach can distinguish the output distribution of a noisy boson sampling  device from that of an approximation  even if the latter  accounts for all multiboson interferences up to order $K\ll \sqrt{N}$ and (ii) the   probability $P_L$ \eqref{Eq5}  essentially depends on multiboson interferences of the  orders above  $K$. These conclusions are very  important in discussion of  the assessments methods  of  boson sampling by verifying  only some low-order correlations at the output distribution,   considered below.

\subsection*{Tests based on   low-order correlators   } 

 The    second-order correlations in  output distribution  were previously used for  assessment of boson sampling    \cite{StatBench,ExpStatSign}.   Current  experiments on boson sampling have  reached  a stage  when the output probability space is  so large that direct verification by comparison with  the output probabilities   $p(\m)$ Eq. \eqref{Eq1}    is out reach  (both computationally and  due to excessively large number of samples for such an assessment). Only the low-order correlations are therefore  checked, such as in the recent benchmark demonstration \cite{20ph60mod}, where only up to $4$th-order correlations have been verified.  Such and other similar assessments are, however,   insufficient, since    some   approximations using a smaller number of interfering bosons, such those of Refs. \cite{R1,RSP,VS2019},   can pass them.  
 
 Let us consider a whole class   of tests    based on  $r$th-order correlations, e.g.,  
\be 
C_r =  \langle \prod_{l=1}^r \hat{b}^\dag_{l} \hat{b}_{l}\rangle -   \prod_{l=1}^r\langle \hat{b}^\dag_{l} \hat{b}_{l}\rangle,
\en{Eq6}   
where  $r=O(1)$.  In   the ideal case of boson sampling with  no noise, for  $N\gg 1$   the   Haar measure,  $\mathrm{Prob}(\ldots)$, of unitary interferometers where  all the occupations $m_1,\ldots, m_M$  of the    output ports   remain bounded by some number  $s$  reads  \cite{Bbirthday}
\be
\mathrm{Prob}( \mathrm{max}(m_l)\le s)  \approx  \left[1-\Bigl(\frac{\rho}{1+\rho}\Bigr)^{s+1}\right]^M. 
\en{Eq7}    
Fixing   $\delta\ll1 $, one can therefore state that  in a randomly selected unitary interferometer, with  high probability  $\mathrm{Prob} \approx 1-\delta$,    the maximal boson bunching count   at  the output  reads   $s =   \ln \left(\frac{N}{\rho\delta}\right)/\ln \left(\frac{1+\rho}{\rho}\right)$.  With the same probability,  the  number of bosons $\mathcal{K}$ detected  in $r$ output ports of the ideal boson sampling scales at most as    $\mathcal{K} = O(r s) = O(r \ln N)$.   This means that with $1-\delta$ chances   any  test based on the $r$-order correlations   will not succeed to tell  the output distribution of   even  the ideal boson sampling   from that of our  classical model with   $K = O(r\ln N)$.  Indeed,  the probability to detect up to $K$ bosons in $r$ output ports    depends on multiboson   interferences only up to the order $K$  \cite{Ninter}.    Due to this fact and the above estimate on the number of bosons detected in $r$ output ports, our model  of Eq. \eqref{JKnew}  with $K = O(r\ln N)$ would  faithfully account for all the correlations  of order $r$, e.g., such as in Eq. \eqref{Eq6},  passing   any test  based  on them. In contrast,  the probability of no counts in a subset of output ports  allows to distinguish  the output distribution produced by such a classical simulation  from that of  the  boson sampling,  with or without  noise,     since the applicability condition of our results, such as in Eq. \eqref{Eq4},   is satisfied: $K = O(r\ln N) \ll \sqrt{N}$ as $N$ scales up.

\section{ Discussion of the results  } 
\label{sec3}

In the spirit of Ref. \cite{F}, we have asked    if a   quantum system realizing  imperfect/noisy  boson sampling  \cite{AA}  can be efficiently and faithfully simulated classically as the system size scales up.  To this goal, we have  investigated  whether it is possible to efficiently distinguish the output distribution of a noisy realization of boson sampling from that   of the  classical approximations  that take into account only low-order multiboson  interferences, where the term ``low-order" means an  order $K\ll \sqrt{N}$, where $N$ is the total number of interfering bosons.  Our  choice of the classical approximations  was dictated by the recent  Refs.   \cite{R1,RSP,VS2019}, which can approximate the output distribution of   boson sampling with any finite noise amplitudes and  to any given error in the total variation distance by adapting the order $K$ of the accounted multiboson interference. 
  
Our  main result is that  one can (and we point  exactly how) efficiently  distinguish noisy boson sampling with finite amplitudes of noise from  the classical approximations  accounting for multiboson  interferences up to an order  $K=O(1)$. The latter  class of  approximation coincides with  the   efficient classical approximations   of Refs.  \cite{R1,RSP,VS2019}.

It is also found that the   required number of runs of a noisy boson sampling for the purpose of distinguishing its output distribution from the considered class of approximations shows  critical dependence    on the scaling with $N$ of the density of bosons $\rho = N/M$,   performing  a transition  from a polynomial in $N$ number of runs   for the  vanishing density of bosons,  e.g.,  in   the   no-collision regime   \cite{AA} (when   $\rho\sim 1/N$ at the least),  to an $N$-independent number of runs  for a finite density of bosons.     

Our results pose some questions. In  the no-collision regime, the lower bound on the total variation distance   between the output distribution of noisy boson sampling   and that of the approximation by low-order multiboson interferences   vanishes when  the number of bosons scales up. On the other hand,  in Refs. \cite{R1,RSP,VS2019}  an upper bound on  the total variation distance  was found, which does not vanish with the total number of bosons.   Are there better efficient classical approximations to the noisy boson sampling, which account for the  multi-boson interferences to the same low order and narrow the gap between the upper and lower bounds? Is it possible to find a better way to distinguish the two output distributions in the no-collision regime, i.e.,  requiring   much smaller  number of samples?  These and related questions are left  for the future work.

 We conclude by noting that the critical dependence on the density of particles,  and not on the number of particles themselves,  is strikingly reminiscent of   the quantum-to-classical   transition   in   systems of    identical particles, occurring  when the  total number of particles   scales up  whereas the    density of  particles  vanishes.  Such a transition results in the mean-field approximation to a large system of identical bosons, the most prominent example being  the Bose-Einstein condensation of weakly interacting bosons \cite{Bogoliubov}, experimentally achieved with dilute gases \cite{BEC1,BEC2} and  approximated   by  the mean-field order parameter obeying the Gross-Pitaevskii equation \cite{Pitaevskii, Gross}.

 \section{Acknowledgements}
 The author acknowledges   correspondence with Scott Aaronson   and discussions with  Jelmer  Renema and Ra\'ul Garc\'ia-Patr\'on. 
 This work   was supported by the National Council for Scientific and Technological Development (CNPq) of Brazil,  Grant 307813/2019-3,  and by the S\~ao Paulo Research Foundation (FAPESP), Grant 2018/24664-9.


\appendix

\section{Lower bound   for  uniformly lossy  interferometer  }
\label{appA}

Here we derive the lower bound on the total variation distance between  the  output    distribution  of  $N$-boson sampling, affected by  boson losses, partial distinguishability of  bosons and dark counts of detectors,     and that of the classical approximation  by  $K$ interfering bosons and $N-K$ classical particles (distinguishable bosons), below called the $K$-reduced model.  Here we  consider the case of  uniformly lossy   interferometer $\mathcal{U} = \sqrt{\eta} U$,  $UU^\dag = I$,   with the transmission $0< \eta\le 1$, whereas  bosons can be in  an arbitrary state of partial distinguishability described by a function $J(\sigma)$ on permutations $\sigma\in S_N$,  defined in the main text.

Let us fix the notations. We will use $\mathcal{U}[k_1,\ldots,k_n|l_1,\ldots,l_n]$ for the  submatrix of $\mathcal{U}$ on the rows $k_1,\ldots,k_n$ (input ports with a boson)  and  a multi-set of columns $l_1,\ldots,l_n$ (the output ports with bosons)   corresponding to occupations $\m = (m_1,\ldots, m_M)$, $|\m| \equiv m_1+\ldots+ m_M=n$ (referred below as the output configuration).

Let us first consider  the case of no  dark counts  of detectors  and  give the probability of  detecting $n$ bosons at the output    of a lossy   boson sampling device. For  uniform  losses,  we can assume boson  losses  to occur    at the input \cite{PRS}, hence, with  the probability $\eta^n(1-\eta)^{N-n}$ only $n$ out of  $N$   bosons are   sent through   interferometer $U$. Then the probability $p_0(\m)$ to detect $n$ bosons in an output  configuration $\m$    reads \cite{VS14,PartDist}
\begin{align}
\label{Aq8}
&\! p_0(\m) = \eta^n(1-\eta)^{N-n} \sum_{\bk}p(\bl|\bk),  \\
& \! p(\bl|\bk) \equiv \frac{1}{\m!} \!\sum_{\sigma_{1,2}\in S_n} \!\! J_{\bk} (\sigma_1\sigma_2^{-1})\!\prod_{\alpha=1}^n \! U^*_{k_{\sigma_1(\alpha)},l_\alpha} U_{k_{\sigma_2(\alpha)},l_\alpha}, \nonumber
\end{align} 
where $\bk =(k_1,\ldots,k_n)$,  a   subset of the input ports $1,\ldots,N$, from which the detected bosons originate,  $p(\bl|\bk)$ is the probability of $n$ partially distinguishable bosons from  input ports $\bk$,  whose state of distinguishability is described by the distinguishability function $J_\bk(\sigma)$,    the output ports $\bl =(l_1,\ldots,l_n)$ correspond  to the configuration $\m$, and  the summation runs over the permutations $\sigma_{1,2}\in S_n$ of $n$     bosons  in the input ports $\bk$.   

Now let us describe the effect of the   dark counts of detectors. Assume  that there are  $\mathcal{N}$ total detector clicks in an output configuration $\s = (s_1,\ldots,s_M)$, thus additionally to $n$ detected bosons there are $\mathcal{N}-n$ dark counts (generally, multiple) corresponding  to  an output configuration $\br=(r_1,\ldots,r_M)$, $|\br|= \mathcal{N}-n$ and $\br\subset \s$ (meaning that  $r_l\le s_l$ for all $l=1,\ldots,M$).  The corresponding  probability $p(\s)$    becomes 
\be
p(\s) = \sum_{n=0}^N \eta^n(1-\eta)^{N-n}\!\!\!\sum_{{\br\subset\s \atop |\br|=\mathcal{N}-n}}\!\!\!\frac{\nu^{\mathcal{N}-n}}{\br!}e^{-M\nu} \!\sum_{\bk} p(\bl|\bk),
\en{Aq9}
where the output ports $\bl=(l_1,\ldots,l_n)$  correspond to the output configuration $\m= \s - \br$, $|\m|= n$. The expression of Eq.~\eqref{Aq9} is  complicated,  however, as we need to  sum  the probabilities in Eq. \eqref{Aq9} corresponding to no counts in a subset of output ports,    the effect of dark counts will be  given by a simple factor.

We consider how close is the output probability distribution of Eq. \eqref{Aq9} to that of  the  $K$-reduced model obtained by replacing  $N-K$ partially distinguishable bosons (from randomly selected inputs)    by  completely distinguishable  ones (i.e., classical particles). As explained in the main text, this model corresponds to the probability distribution $p^{(K)}(\m)$ similar to that of Eq. \eqref{Aq9} but with the   following  distinguishability function  
   \be
J^{(K)}(\sigma) \equiv \left\{\begin{array}{cc}J(\sigma), & c_1(\sigma) \ge N-K,\\ 0, & c_1(\sigma)< N-K.  \end{array}   \right.
\en{Aq10}
Consider a lower bound  on the total variation distance $\mathcal{D}=\mathcal{D}(p,p^{(K)})$ between our  noisy  boson sampling and its   $K$-reduced model.   We have $\mathcal{D} = \sum_n \mathcal{D}_n$, where $\mathcal{D}_n$ corresponds to exactly $n$ detected bosons at the output.  The lower bound is given by the absolute value of the difference in  probability to detect all the output bosons     in a fixed  subset of   output ports, below set to be   $L+1\le l\le  M$, i.e.,  no counts  in output ports $\Omega_L = \{1, \ldots, L\}$.  In this case, the effect of dark counts can be accounted  in a very simple manner. Being  independent of the input state of bosons,    they contribute   the  factor $e^{-L\nu}$  to the probability  of zero counts  in a subset of $L$ output ports.  Therefore, we have  
\begin{align}
\label{Aq11_OLD}
 &  \mathcal{D} =\sum_{n=0}^N \mathcal{D}_n \ge \left| P_L -P^{(K)}_L\right|, \\
 &   P_L \equiv e^{-\nu L} \sum_{n=0}^N { \sum_{|\m|=n}}^\prime  p_0(\m),\nonumber\\
 & P^{(K)}_L \equiv  e^{-\nu L} \sum_{n=0}^N{\sum_{|\m|=n}}^\prime  p^{(K)}_0(\m), \nonumber 
   \end{align} 
  where the summations   with prime run   over all output   configurations $\m=(0,\ldots,0,m_{L+1},\ldots, m_M)$ with no counts in the first $L$  output ports.  Denoting $\Delta P_L = P_L -P^{(K)}_L$ (and similar for other quantities of the two models, e.g. $\Delta J = J - J^{(K)}$, etc) and  using  
  Eq.~\eqref{Aq8} after some algebra we obtain  
  \begin{align}
\label{Aq13}
& \!\Delta P_L   =   e^{-\nu L}\sum_{n=0}^N \eta^n(1-\eta)^{N-n}\sum_{\bk}{\sum_{|\m|=n}}^\prime\Delta p(\bl|\bk)
\nonumber\\
&= \! e^{-\nu L} \sum_{n=0}^N \eta^n(1-\eta)^{N-n}\sum_{\bk} \!\!\sum_{\sigma\in S_n} \Delta J_\bk(\sigma)\!\prod_{\alpha=1}^n H_{k_\alpha, k_{\sigma(\alpha)}}  \nonumber\\
&=\!  e^{-\nu L}\sum_{\tau\in S_N} \Delta J (\tau)\prod_{k=1}^N A_{k,\tau(k)} ,
\end{align}
where  $H$ and $A$ are   $N$-dimensional positive  semidefinite Hermitian matrices, defined as follows: 
\begin{align}
 \label{Aq12}
& H_{kj} = \sum_{l=L+1}^M U_{kl} U^*_{jl}, \\
& A_{kj} = (1-\eta)\delta_{kj} + \eta H_{kj} = \delta_{kj}- \eta\sum_{l=1}^LU_{kl}U^*_{jl}.  \nonumber
\end{align}
To arrive  at the result in Eqs. \eqref{Aq13} and \eqref{Aq12} we have  performed the following  steps.  We have used an identity for the sum of  output  probabilities     \cite{BB,VS2017}
\begin{align*}
&{\sum_{|\m|=n}}^\prime \Delta p(\bl|\bk) =\sum_{l_1=L+1}^M \ldots \sum_{l_n=L+1}^M  \frac{\m!}{n!} \Delta p(\bl|\bk)\\
& = \frac{1}{n!} \sum_{\sigma_{1,2}\in S_n} \Delta J_\bk(\sigma_1\sigma^{-1}_2) \prod_{\alpha=1}^n H_{k_{\sigma_2(\alpha)},k_{\sigma_1(\alpha)}}\\
& = \frac{1}{n!} \sum_{\sigma_{1,2}\in S_n} \Delta J_\bk(\sigma_1\sigma^{-1}_2) \prod_{\alpha=1}^n H_{k_\alpha,k_{\sigma_1\sigma_2^{-1}(\alpha)}}\\
& = \sum_{\sigma\in S_n} \Delta J_\bk(\sigma) \prod_{\alpha=1}^n H_{k_\alpha,k_{\sigma(\alpha)}},
\end{align*}
where     $\sigma \equiv \sigma_1\sigma^{-1}_2$,  the following  identity 
\be
 \prod_{\alpha=1}^n H_{k_{\sigma_2(\alpha)},k_{\sigma_1(\alpha)}}=  \prod_{\alpha=1}^n H_{k_\alpha,k_{\sigma_1\sigma_2^{-1}(\alpha)}},
 \en{IdP}
replaced the  sum over all output configurations $\m$ by  $n$ independent sums  over   output ports $l_1,\ldots,l_n$, using that  for  any symmetric function $f(l_1,\ldots,l_n)$,   
\be
\sum_{ |\m|=n} f(l_1,\ldots,l_n) =\!\! \sum_{l_1=L+1}^M \!\!\ldots \!\!\sum_{l_n=L+1}^M  \frac{\m!}{n!}f(l_1,\ldots,l_n),
\en{SumId}
   and     observed that for any permutation $\tau \in S_N$ 
\begin{align}
\label{Prod}
 & \prod_{k=1}^N A_{k,\tau(k)} = \sum_{n=0}^N \eta^n(1-\eta)^{N-n} \nonumber\\
 &\times \sum_{\bk} \prod_{\alpha=1}^n H_{k_\alpha,\tau(k_\alpha)}  \prod_{\alpha = n+1}^N \delta_{k_\alpha,\tau(k_\alpha)} \nonumber\\
& = \sum_{n=0}^N \eta^n(1-\eta)^{N-n} \sum_{\bk} \prod_{\alpha=1}^n 
H_{k_\alpha, k_\sigma(\alpha)},
\end{align}
where    $\sigma\in S_n$, acting on the subindex in $k_\alpha$,   is defined   by the relation  $k_{\sigma(\alpha)} = \tau(k_\alpha)$  (we can introduce such $\sigma$, since  $\tau$ permutes  elements in  $\bk$ between themselves).

  To derive Eq. \eqref{Aq13} we have used here that  the  interferometer has  uniform transmission  $\eta$. However,  as proven below,  the resulting expression   applies for an arbitrary (non-uniformly)  lossy interferometer $\mathcal{U}$. 

\medskip 
\section{Lower bound for  arbitrary lossy interferometer    } 
\label{appB}

Recall that an arbitrary  lossy linear interferometer  with  $M$ input ports can be imbedded into a unitary one with  $2M$ input ports \cite{LossNet} with introduction of $M$ auxiliary boson modes describing losses.   One such  $2M$-dimensional unitary interferometer reads  \cite{BB}
\be
\hat{\mathcal{U}} = \left( \begin{array}{cc} \mathcal{U} & V \\ - V^\dag F &\sqrt{ D}
\end{array} \right), 
\en{B1} 
where  $\mathcal{U} = \sqrt{B} F $,  with  $B = \mathcal{U}\mathcal{U}^\dag$ and a unitary matrix  $F$,  $V$ is such that $VV^\dag = I - \mathcal{U} \mathcal{U}^\dag$, and $D = \mathrm{diag}(\eta_1,\ldots, \eta_M)$ with $\eta_l$ being  eigenvalue of $B$ (a singular value of  $\mathcal{U}$). 
Bosons at the output ports $M+1\le l\le 2M$ correspond to losses. Below we will only need to consider the $M\times 2M$-dimensional block $(\mathcal{U}, V)$.  For zero dark counts, the probability that $n$ out of input $N$ partially distinguishable bosons are detected at  the output  of a lossy interferometer $\mathcal{U}$ in an output configuration $\m = (m_1,\ldots,m_M)$ can be obtained by  summation of the general formula for   output probability     $p(\m)$ in the  unitary interferometer $\hat{\mathcal{U}}$  \cite{PartDist}, similar as in Eq. \eqref{Aq8}, over all possible output configurations $\br = (r_1, \ldots, r_M)$, $|\br| = N-n$,   of bosons in  the output  ports  $M+1 \le l \le 2M$. Since the probability is symmetric in the input/output ports, we designate output ports $l_1,\ldots,l_n$ to be the detected bosons, i.e., $1\le l_k \le M$, $1\le k\le n$. Using  a summation identity similar to that of Eq. \eqref{SumId},
\begin{align}
\label{IdS}
& \sum_{|\br|=N-n}f(l_1,\ldots,l_N) \nonumber\\
& = \sum_{l_{n+1}=M+1}^{2M} \ldots  \sum_{l_{N}=M+1}^{2M} \frac{\br!}{(N-n)!}f(l_1,\ldots,l_N),\nonumber\\
\end{align} 
and recalling that $VV^\dag = I- \mathcal{U}\mathcal{U}^\dag$ we obtain 
\begin{widetext}
\begin{align}
\label{B2}
 p(\m) &= \sum_{|\br|=N-n}\frac{1}{\m!\br!} \sum_{\sigma_{1,2}\in S_N} J(\sigma_1\sigma_2^{-1})
\left[ \prod_{k=1}^n \mathcal{U}^*_{\sigma_1(k),l_k} \mathcal{U}_{\sigma_2(k),l_k} \right]\prod_{k=n+1}^N V^*_{\sigma_1(k),l_k} V_{\sigma_2(k),l_k}\nonumber\\
  &= \frac{1}{(N-n)!\m!}\sum_{\sigma_{1,2}\in S_N} J(\sigma_1\sigma_2^{-1})
\left[ \prod_{k=1}^n \mathcal{U}^*_{\sigma_1(k),l_k} \mathcal{U}_{\sigma_2(k),l_k} \right]\prod_{k=n+1}^N(I- \mathcal{U}\mathcal{U}^\dag)_{\sigma_2(k),\sigma_1(k)}\nonumber\\
& = \frac{1}{\m!}\sum_{\sigma_R\in S_N} J(\sigma_R)\sum_{\tau\in S_n}\sum_{\bk}
\left[ \prod_{\alpha=1}^n \mathcal{U}^*_{\sigma_R\tau(k_\alpha),l_\alpha} \mathcal{U}_{\tau(k_\alpha),l_\alpha} \right]\prod_{\alpha=n+1}^N(I- \mathcal{U}\mathcal{U}^\dag)_{k_\alpha, \sigma_R(k_\alpha)},\nonumber\\
\end{align}
\end{widetext}
where we have introduced  subsets  $\bk = (k_1,\ldots,k_n)$ of the  input ports   of $1,\ldots, N$, where    $k_\alpha \equiv \sigma_2(\alpha)$, used an identity similar to that of Eq. \eqref{IdP}
to recast    the summation over $\sigma_2\in S_N$ as that over  the subsets $\bk$  and permutations $\tau\in S_n$ in each such subset,  and introduced the relative permutation $\sigma_R = \sigma_1\sigma^{-1}_2$. Note that the r.h.s. of Eq. \eqref{B2} does not depend on the  elements of the auxiliary unitary matrix $\hat{\mathcal{U}}$ other than $\mathcal{U}$.

Consider now  the probability $P_L(n)$ of no counts in $L$ output ports   $l = 1,\ldots, L$  with  exactly $n$ bosons  detected at the output of $\U$. For zero dark counts, by using the  summation  identity of   Eq. \eqref{SumId} we get
\begin{widetext}
\begin{align}
\label{B3}
 P_L(n) &= \sum_{l_1=L+1}^{M} \ldots \sum_{l_n=L+1}^{M} \frac{\m!}{n!} p(\m) = \frac{1}{n!} \sum_{\sigma\in S_N} J(\sigma)\sum_{\tau\in S_n}\sum_{\bk}  \left[\prod_{\alpha=1}^n  \mathcal{H}_{\sigma\tau(k_\alpha),\tau(k_\alpha)} \right] \prod_{\alpha=n+1}^N (I- \mathcal{U}\mathcal{U}^\dag)_{\sigma(k_\alpha),k_\alpha}\nonumber\\
& = \sum_{\sigma\in S_N} J(\sigma)\sum_{\bk}  \left[\prod_{\alpha=1}^n \mathcal{H}_{\sigma(k_\alpha),k_\alpha} \right] \prod_{\alpha=n+1}^N (I- \mathcal{U}\mathcal{U}^\dag)_{\sigma(k_\alpha),k_\alpha},
\end{align}
\end{widetext}
where  we have introduced a positive semi-definite Hermitian matrix 
\be
 \mathcal{H}_{kj} = \sum_{l=L+1}^{M} \mathcal{U}_{kl}\mathcal{U}^*_{jl}
\en{matH}
and used an identity for  permutation  $\tau$, similar as in Eq. \eqref{IdP}. 

Thanks to the    summation  in   Eq. \eqref{B3} over all  products of matrix elements on  exactly  $n$ input ports $\bk = (k_1,\ldots, k_n)$  selected from  the total $ N$ of them, the result of Eq. \eqref{B3}  can be    rewritten  as the $n$th order derivative in  an auxiliary variable $x$ at $x=0$ of the    $N$th order polynomial:
\be
\!\!\! P_L(n) \!= \!\frac{1}{n!} \frac{d^n}{dx^n}
 \sum_{\sigma\in S_N} J(\sigma)\prod_{k=1}^N\left(x  \mathcal{H} \!+\! I\!-\! \mathcal{U}\mathcal{U}^\dag\right)_{k,\sigma(k)}.
\en{B4}
Eq. \eqref{B4} is very convenient for the derivation of the expression for  the total probability of no counts in $L$ output ports $P_L = \sum_{n=0}^NP_L(n)$. Indeed, since   the polynomial in Eq. \eqref{B4} is of  order $N$,  the latter  sum is  the Taylor series  about $x=0$  with $\Delta x  =1$,  with the result being  the  value of the polynomial  at $x=1$.  Taking into  account  also  the random counts   factor $e^{-\nu L}$, we get   
 \be
 P_L = e^{-\nu L}\sum_{\sigma\in S_N} J(\sigma)\prod_{k=1}^N\left(  \mathcal{H} + I- \mathcal{U}\mathcal{U}^\dag\right)_{k,\sigma(k)}.
 \en{B5}
Finally, introducing  a positive semidefinite Hermitian matrix $A$, similar as in Eq. \eqref{Aq12}, we obtain from Eq. \eqref{B5}:
\begin{align}
\label{B6}
& \Delta P_L = e^{-L\nu}   \sum_{\sigma\in S_N} \Delta J(\sigma)\prod_{k=1}^NA_{k,\sigma(k)},\nonumber \\
&  A_{kj} \equiv  \mathcal{H}_{kj} + \delta_{kj}- \sum_{l=1}^M\mathcal{U}_{kl}\mathcal{U}^*_{jl}= \delta_{kj} - \sum_{l=1}^L\mathcal{U}_{kl}\mathcal{U}^*_{jl},\nonumber\\
\end{align}
where  $\Delta J = J - J^{(K)}$.

\section{ Estimating  $\Delta P_1$}
\label{appC}

Here we derive an analytical estimate on the difference in probability in  Eq. \eqref{B6}  for  $L=1$  in the simplest case of a uniform partial distinguishability $\xi$    and a uniform transmission  $\eta$,  $\mathcal{U} = \sqrt{\eta} U$.  In this case, as explained in the main text, the distinguishability function  for $N$ bosons reads  $J(\sigma) = \xi^{N-c_1(\sigma)}$, where  $c_1(\sigma)$    is the number of fixed points in  permutation $\sigma$.   Setting    the output port with no counts  to be $l=1$ and expanding the product in Eq. \eqref{B6}  by using the expression   $A_{kj}= \delta_{kj}- \eta U_{k1}U^*_{j1}$,     we obtain  
\begin{align}
\label{Aq15}
\Delta P_1& =  e^{-\nu }   \sum_{\tau\in S_N} \Delta J (\tau)\prod_{k=1}^N \left( \delta_{k,\tau(k)} - \eta U_{k,1}U^*_{\tau(k),1} \right)  \nonumber\\
&= e^{-\nu }  \sum_{n=0}^N (-\eta)^n \sum_{\sigma\in S_n} \Delta J_n (\sigma) \sum_\bk \prod_{\alpha=1}^n |U_{k_\alpha,1}|^2 \nonumber\\
\end{align}
where  we have partitioned the set $1,\ldots, N$ into two subsets $\bk = (k_1,\ldots,k_n)$ and $\bk^\prime=(k_{n+1},\ldots, k_{N})$, such that    $\bk^\prime $ contains the  fixed points of $\tau$,   taken into  account that $\prod_{\alpha=1}^n U_{k_\alpha,1}U^*_{\tau(k_\alpha),1} = \prod_{\alpha=1}^n |U_{k_\alpha,1}|^2$,  introduced  permutation $\sigma \in S_n$ such that $\tau(k_\alpha)= k_{\sigma(\alpha)}$ for $\alpha=1,\ldots, n$, and   used  that $J(\tau)  = \xi^{n-c_1(\sigma)}\equiv J_n(\sigma)$ for    $\tau= I\otimes \sigma $ with $\sigma\in S_n$ and $I\in S_{N-n}$ (the identity permutation), in this case  $c_1(\tau) = N-n+c_1(\sigma)$. By Eq. \eqref{Aq10} $\Delta J_n(\sigma)=J_n(\sigma)- J^{(K)}_n(\sigma)$ becomes
\be
\Delta J_n(\sigma)  = \left\{\begin{array}{cc}0 & c_1(\sigma) \ge n-K,\\ \xi^{n-c_1(\sigma)}, & c_1(\sigma)< n-K.  \end{array}   \right.
\en{Aq16}
Since $\Delta J_n(\sigma)$ depends only on the number of fixed points $c_1(\sigma)$, we can perform summation  in Eq. \eqref{Aq15} over the permutations 
\be
\sum_{\sigma\in S_n} \Delta J_n (\sigma) =  \sum_{m=K+1}^{n}\binom{n}{m}d_{m} \xi^{m} ,
\en{Aq17}
where we have used an identity for   permutations in $S_n$  having  $n-m$ fixed points \cite{Stanley} (see also Eq. \eqref{F4} in appendix \ref{appF})
\be
 \sum_{\sigma\in S_n}\delta_{c_1(\sigma),n-m} = \binom{n}{m} d_{m}, \quad d_{m} \equiv m! \sum_{s=0}^{m}\frac{(-1)^s}{s!}.
\en{Aq18}
Using  Eq. \eqref{Aq17} into Eq. \eqref{Aq15} and observing that by Eq. \eqref{Aq16} nonzero terms have  $n\ge K+1$ we obtain   
\begin{align}
 \Delta P_1  &= e^{-\nu }  \sum_{n=K+1}^N (-\eta)^n \sum_{m=K+1}^{n}\binom{n}{m}d_{m} \xi^{m}  \nonumber \\
 &\times \sum_\bk \prod_{\alpha=1}^n |U_{k_\alpha,1}|^2.
\label{Aq19}
\end{align}

\subsection{Balanced output port: lower bound on  $\left|\Delta P_1\right|$}
\label{subsec1}

Let us first estimate the r.h.s. of Eq. \eqref{Aq19} for the interferometer with a balanced output port $|U_{k,1}|^2 = 1/M$. In this case we can simplify  the sum in Eq. \eqref{Aq19}  as follows 
\begin{align}
\label{Aq20}
&  \sum_{n=K+1}^N (-\eta)^n \sum_{m=K+1}^{n}\binom{n}{m}d_{m} \xi^{m}  \sum_\bk \frac{1}{M^n} \nonumber\\
&  = \sum_{n=K+1}^N \binom{N}{n} \left(-\frac{\eta}{M}\right)^n \sum_{m=K+1}^{n}\binom{n}{m}d_{m} \xi^{m}\nonumber\\
&  = \sum_{m=K+1}^N d_m \xi^m\sum_{n=m}^N \binom{N}{n}\binom{n}{m}\left(-\frac{\eta}{M}\right)^n \nonumber\\
&  = \sum_{m=K+1}^N d_m  \binom{N}{m}  \left(-\frac{\xi\eta}{M}\right)^m \left( 1- \frac{\eta}{M}\right)^{N-m},
\end{align}
where we have  used the binomial  identity in the sum over $n-m$.  We  have from Eqs. \eqref{Aq19} and \eqref{Aq20}
\begin{align}
\label{Aq21}
 \Delta P_1 =     e^{ -\nu } \left( 1- \frac{\eta}{M}\right)^{N} \!\! \sum_{m=K+1}^N  \binom{N}{m}   d_m  \left(-X\right)^m ,
\end{align}
where we have introduced 
\be
X \equiv  \frac{\xi\eta }{M}\left( 1- \frac{\eta}{M}\right)^{-1} = \frac{\rho\xi\eta}{N- \rho\eta} .
\en{Aq22}
Using  the Pochhammer symbol for falling factorial, $(n)_m = n(n-1)\ldots (n-m+1)$,   the  sum on the r.h.s. of Eq. \eqref{Aq21} can be rewritten  as follows
\begin{align}
 \label{Aq23}
&&\sum_{m=K+1}^N   \binom{N}{m}   d_m  (-X)^m  = (N)_{K+1} (-X)^{K+1} \nonumber\\
&& \times \sum_{s=0}^{N-K-1}  (N-K-1)_s (-X)^s\frac{d_{K+1+s}}{(K+1+s)!}\nonumber\\
&& \equiv  (N)_{K+1} (-X)^{K+1} I_{N-K-1}(K).
\end{align}
Let us  consider the remaining sum in Eq. (\ref{Aq23}), denoted by $I_{N-K-1}(K)$.  Using the estimate on $d_m$ in Eqs. (\ref{F4})  and (\ref{R_m}) obtained in  Section \ref{appF}, we get
\begin{eqnarray}
\label{Aq24} 
&& I_n(K)= \sum_{s=0}^n  (n)_s (-X)^s\frac{d_{K+1+s}}{(K+1+s)!}\nonumber\\
&&=e^{-1}\sum_{s=0}^n  (n)_s (-X)^s \left[1 + O\left(\frac{1}{(K+1+s)!}\right)\right]\nonumber\\
&&=e^{-1}\sum_{s=0}^n  (n)_s (-X)^s \left[1 + O\left(\frac{1}{(K+1)!}\right)\right]\nonumber\\
&& \ge  \frac{e^{-1}}{1+ nX}\left[ 1+ O\left(\frac{1}{(K+1)!}\right)\right],\nonumber\\
\end{eqnarray}
where  we have used that  $nX<1$, more precisely $nX = (N-K-1)X = \rho \xi \eta(1+O(K/N))$, hence  
\begin{eqnarray}
\label{Aq25}
&& \sum_{s=0}^n  (n)_s (-X)^s = 1 - nX(1-[n-1]X) \nonumber\\
&&  - n(n-1)(n-2)X^3(1- [n-3]X) - \ldots   \nonumber\\
&& \ge \sum_{s=0}^n   (-nX)^s + O\left[(nX)^n\right]\nonumber\\
&& =  \frac{1}{1+ nX}+O\left[(nX)^n\right].
\end{eqnarray} 
The exponentially small term in Eq. \eqref{Aq25} can be dropped  in Eq. \eqref{Aq24}  as compared to the  term $\frac{1}{(K+1)!}$.  
 
What is left is to estimate the front factors in Eq. \eqref{Aq21} and \eqref{Aq23}. We have 
\be
\left(1- \frac{\eta}{M}\right)^N = \exp(-\rho\eta)\left[1- \left|O\left(\frac{1}{N}\right)\right|\right].
\en{Aq28}
Using the definition Eq. \eqref{Aq22} and the following identity (derived by using Euler's summation formula, similarly as in  Ref. \cite{VS2017})
\[
\prod_{l=1}^K\left(1-\frac{l}{N}\right) = \exp\!\left(-\frac{K(K+1)}{2N}\right)\!\left(\!1 \!+\!O\left( \frac{K^3}{N^2}\right)\! \right),
\]
we obtain 
\begin{align}
 \label{Aq29}
 &  (N)_{K+1} X^{K+1}=  \left(\frac{\rho \xi\eta}{1-\rho\eta/N} \right)^{K+1} \prod_{l=1}^K \left(1- \frac{l}{N}\right) \nonumber\\
 &  = \left(\rho \xi\eta e^{-\frac{K/2+\xi\eta}{N}}\right)^{K+1}\left[1+O\left(\frac{K^3}{N^2}\right)\right]\nonumber\\
 & =  (\rho \xi\eta )^{K+1}\left[1- \left|O\left(\frac{K^2}{N}\right)\right|\right].
\end{align} 
 
    Combining    Eqs. \eqref{Aq21}  -  \eqref{Aq29}   we get the following  result
\begin{align}
\label{Aq30}
  (-1)^{K+1}\Delta P_1 &\ge   W_1 \left[1- \left|O\left( \frac{K^2}{N}+   \frac{1}{(K+1)!}\right)\right|\right] \\
  W_1 &\equiv  \frac{(\eta\xi \rho)^{K+1}}{1+\eta\xi \rho} e^{- 1 -\nu -\eta \rho}.\nonumber
 \end{align}

 Below  will show  that the leading order of $W_1$ from   Eq.~\eqref{Aq30}  is also the leading order of the lower bound    in  a Haar-random interferometer under a similar  condition $K^2\ll N$, whereas the variance scales  as $O(K^2/N)$.  
   
\subsection{The average $\langle \Delta P_1\rangle $ in the  Haar-random interferometer  }
\label{subsec2}

We will use the following known average in  a   Haar-random unitary interferometer $U$    \cite{HaarR}
 \be
 \langle \prod_{\alpha=1}^n |U_{k_\alpha,1}|^{2s_k} \rangle =    \frac{s_1!\ldots s_n!}{M^{(n)}},
\en{Eav}
where  $m^{(n)} \equiv m(m+1)\ldots (m+n-1)$ (Pochhammer symbol for    the rising factorial).  

Consider the average lower bound  in a Haar-random interferometer obtained by averaging  Eq. \eqref{Aq19}. Observing that $\langle |x|\rangle \ge |\langle x\rangle|$, we have $\langle \mathcal{D}\rangle \ge \left| \langle \Delta P_1\rangle \right|$. We obtain 
\begin{align}
 \label{Aq31}
 \!\!\langle  \Delta P_1 \rangle &= e^{-\nu}  \!\!\sum_{n=K+1}^N (-\eta)^n  \!\!\sum_{m=K+1}^{n}\binom{n}{m}d_{m} \xi^{m}  \! \sum_\bk \frac{1}{M^{(n)}}  \nonumber\\
\!\! & =  e^{-\nu}  \!\! \sum_{n=K+1}^N \binom{N}{n}  \frac{(-\eta)^n}{M^{(n)}}  \!\!\sum_{m=K+1}^{n}\binom{n}{m}d_{m} \xi^{m} ,\nonumber\\
\end{align}
where the only difference from the similar expression in Eq. \eqref{Aq20} is the Pochhammer symbol $M^{(n)}$ instead of $M^n$ in the denominator. The factors in the first sum in Eq. \eqref{Aq31} have the following upper bound
\be
\left|\binom{N}{n}  \frac{(-\eta)^n}{M^{(n)}} \right| < \frac{(\rho \eta)^n}{n!},
\en{Boundtn} 
thus they decrease exponentially fast with $n$.  Due to  the expansion  (derived by using Euler's summation, similar as in  Ref.    \cite{VS2017})
\be
M^{(n)} = M^n e^{ \frac{n^2}{2M}}\left(1 +O\left(\frac{n^3}{M^2}\right) \right), 
\en{Poch_1}
we can neglect the    difference between $(\eta/M)^n$ and $\eta^n/M^{(n)}$ for  $n^2\ll {M}$. These two observations allows us to conclude that, without much of error,  one can substitute the Pochhammer symbol $M^{(n)}$ by $M^n$ on the r.h.s. of Eq. \eqref{Aq31}, thus obtaining  the same estimate on the average lower bound as in Eq. \eqref{Aq30}, if  for the few first terms  in the sum   $n^2  \ll M  = N/\rho$, which necessitates that $K^2\ll N/\rho$.  Below  we give an alternative derivation of  the   leading order  of the average lower bound in Eq. \eqref{Aq31}. 

 Using  Pochhammer symbol  for  the falling factorial,  $(m)_n \equiv m(m-1)\ldots (m-n+1)$, we get
\begin{align}
\label{Aq32}
& \sum_{n=K+1}^N \binom{N}{n}  \frac{(-\eta)^n}{M^{(n)}} \sum_{m=K+1}^{n}\binom{n}{m}d_{m} \xi^{m} \nonumber\\
&  = \sum_{n=K+1}^N \frac{(N)_n}{ M^{(n)}} (-\eta)^n \sum_{m=K+1}^n \frac{d_m}{m!} \frac{\xi^m}{(n-m)!}
\nonumber\\
&  = \frac{(N)_{K+1}}{(M)_{K+1}}(-\eta\xi)^{K+1}\sum_{l=0}^{N-K-1}\frac{(N-K-1)_l}{(M+K+1)_l}(-\eta)^l   \nonumber\\
&  \times \sum_{s=0}^l \frac{d_{s+K+1}}{(s+K+1)!}\frac{\xi^s}{(l-s)!},
\end{align}
where we have introduced new indices $l$ and $s$ by  $n = l+K+1$ and $m = s+K+1$.  Due to the above discussion, we can  approximate  the falling factorial in the  reduced sum by the corresponding power, using that   (obtained similar as  Eq. \eqref{Poch_1})
\be
(N)_n = N^n e^{ -\frac{n^2}{2N}}\left(1 +O\left(\frac{n^3}{N^2}\right) \right).
\en{Poch_2}
Assuming that $K$ is not small for the  approximation  $d_m/m! \approx  e^{-1} $  for $m\ge K+1$ (see Eq. \eqref{Aq18}) and further details in  Eq. \eqref{F4} of  appendix \ref{appF})  and  approximating the exponential series by  the exponential functions for \mbox{$N-K\gg 1$}  we have 
 \begin{align} 
 \label{Aq33}
&\! \sum_{l=0}^{N-K-1} \frac{(N-K-1)_l}{(M+K+1)_l}(-\eta)^l \sum_{s=0}^l \frac{d_{s+K+1}}{(s+K+1)!}\frac{\xi^s}{(l-s)!}\nonumber\\
& \! \approx e^{-1} \sum_{l=0}^{N-K-1} \left( -\frac{(N-K-1)\eta}{M+K+1}\right)^l\sum_{s=0}^l \frac{\xi^s}{(l-s)!}\nonumber \\
 &\!  = e^{-1}\sum_{s=0}^{N-K-1} \xi^s \sum_{l=s}^{N-K-1} \frac{1}{(l-s)!} \left( -\frac{(N-K-1)\eta}{M+K+1}\right)^l \nonumber\\
 &\!  = e^{-1}\sum_{s=0}^{N-K-1} \left( -\frac{(N-K-1)\xi\eta}{M+K+1}\right)^{\! s} \sum_{l=s}^{N-K-1} \frac{1}{(l-s)!} \nonumber\\
 &\! \times \left( -\frac{(N-K-1)\eta}{M+K+1}\right)^{l-s}
 \nonumber\\
 & \!\!\approx\!\!  \sum_{s=0}^{N-K-1}\!\! \left( -\frac{(N\!-\!K\!-\!1)\xi\eta}{M\!+\!K\!+\!1}\right)^s \exp\left\{\!-1\!-\!\frac{N\!-\!K\!-\!1}{M\!+\!K\!+\!1}\eta\right\}\nonumber\\
& \approx\frac{\exp\{-1-\frac{N-K-1}{M+K+1} \eta \}}{1 + \frac{N-K-1}{M+K+1}\eta \xi} \nonumber\\
&  = \frac{\exp\{-1- \rho \eta \}}{1 + \rho\eta \xi}\left[1+ O\left(\frac{K}{N}\right)\right].\nonumber\\
\end{align}
In the same way, we  can also  approximate  the ratio of two falling factorials in the first  factor in Eq. \eqref{Aq32}
\[
 \frac{(N)_{K+1}}{(M)_{K+1}}(-\eta\xi)^{K+1}  =(-\rho\xi\eta)^{K+1}\left[ 1 + O\left(\frac{K^2}{N}\right)\right].
\]
 Therefore,  we obtain from Eqs. \eqref{Aq31}, \eqref{Aq32} and \eqref{Aq33} 
\be
\langle \Delta P_1 \rangle \approx   \frac{(-\eta\xi \rho)^{K+1}}{1+\eta\xi \rho} e^{- 1 -\nu -\eta \rho}
\en{Aq34}
where now the lower bound of Eq. \eqref{Aq30} is an approximation to the average for  sufficiently large values of $K$ and $N\gg K^2$. 

Below, using the same  approximations as those used for derivation of Eq. \eqref{Aq34}, we show that  the relative variance  of  the difference in the probability in Eq. \eqref{Aq19} is of the  order $O(K^2/N)$, which means that except for a vanishing fraction of interferometers as $N$ scales up, the average   result in  Eq. \eqref{Aq34} is the leading order of an absolute  lower bound.

\subsection{Variance of $\Delta P_1$ in the  Haar-random interferometers}
\label{subsec3}

Consider the variance of  $\Delta P_1 $ in Eq. \eqref{Aq19}.    Due to     Eq. \eqref{Eav}    the  variance   $\mathcal{R} = \langle(\Delta P_1)^2\rangle - \langle\Delta P_1 \rangle^2$  becomes 
 \begin{align}
&   \mathcal{R} = e^{-2\nu}\sum_{n_1=K+1}^N  \sum_{n_2=K+1}^N (-\eta)^{n_1+n_2}  \nonumber\\
&  \!\times\!\! \sum_{m_1=K+1}^{n_1}\binom{n_1}{m_1}d_{m_1} \xi^{m_1}\!\!\!\!\sum_{m_2=K+1}^{n_2}\!\!\binom{n_2}{m_2}d_{m_2} \xi^{m_2}  \Theta_{n_1,n_2},
\nonumber\\
\label{Aq47}
\end{align}
with 
\begin{align}
 &   \Theta_{n_1,n_2} \equiv \sum_{\bk^{(1)}}  \sum_{\bk^{(2)}}  \langle \prod_{i=1,2} \prod_{k\in \bk^{(i)} } |U_{k,1}|^{2}  \rangle \nonumber\\
 &  -\sum_{\bk^{(1)}}  \sum_{\bk^{(2)}}\prod_{i=1,2} \langle \prod_{k\in \bk^{(i)} } |U_{k,1}|^{2}  \rangle  \nonumber\\
 &  = \sum_{\bk^{(1)}}  \sum_{\bk^{(2)}}   \left[\frac{2^{|\bk^{(1)}\cap\bk^{(2)}|}}{M^{(n_1+n_1)}} - \frac{1}{M^{(n_1)}} \frac{1}{M^{(n_2)}} \right],
 \label{Aq48}
\end{align}
 where we have used   that      for a common  element  $k$ in $\bk^{(1,2)}$  (denoting  this subset by $ \bk^{(1)}\cap\bk^{(2)}$) there is a factor $2$ in the r.h.s. of Eq. \eqref{Eav}. 
Let us simplify the   sum  in Eq.~\eqref{Aq48}. By assuming that $n_2\ge n_1$ and using a dummy variable $a\in \{1, 2\}$ we get for the  two terms in the square brackets in Eq. \eqref{Aq48}
\begin{align}
& \! \sum_{\bk^{(1)}}  \sum_{\bk^{(2)}} a^{|\bk^{(1)}\cap\bk^{(2)}|} = \binom{N}{n_2}\sum_{s=0}^{n_1} \binom{n_2}{s}\binom{N-n_2}{n_1-s} a^s \nonumber \\
& \!\! = \!\!\sum_{s=0}^{\mathrm{min}(n_1,n_2)}  \frac{a^s}{s!(n_1-s)!(n_2-s)!}\frac{N!}{(N-n_1-n_2+s)!}.\nonumber\\
\label{AqE}\end{align}
Since the   expression in Eq. \eqref{AqE} is symmetric  in $n_1$ and $n_2$, it  applies  for any $n_1$ and $n_2$. Now,    using the Pochhammer notations  for the falling factorials, for $ \theta_{n_1,n_2} \equiv n_1! n_2! \Theta_{n_1,n_2}$ we get
\begin{align}
\label{Aq49}
  \theta_{n_1,n_2} &=  \sum_{s=0}^{\mathrm{min}(n_1,n_2)}  \frac{(n_1)_s (n_2)_s (N)_{n_1+n_2-s}}{s! M^{(n_1+n_2)}}\nonumber\\
 &  \times \left(2^s - \frac{M^{(n_1+n_2)}}{M^{(n_1)} M^{(n_2)}}\right). 
 \end{align}
By setting $n_i = l_i +K+1$ and $m_i = s_i+K+1$ in Eq. \eqref{Aq47}   we obtain using Eq. \eqref{Aq49}
 \begin{align}
\label{Aq50} 
&   \quad\mathcal{R} = e^{-2\nu}\sum_{n_1=K+1}^N  \sum_{n_2=K+1}^N (-\eta)^{n_1+n_2} \theta_{n_1,n_2}\nonumber\\
& \quad \times \!\!\! \sum_{m_1=K+1}^{n_1}\frac{d_{m_1}}{m_1!} \frac{\xi^{m_1}}{(n_1-m_1)!}\sum_{m_2=K+1}^{n_2}\frac{d_{m_2}}{m_2!} \frac{\xi^{m_2}}{(n_2-m_2)!}\nonumber\\
 &   \quad=(\eta \xi )^{2K+2}e^{-2\nu}\sum_{l_1,l_2=0}^{N-K-1}  (-\eta)^{l_1+l_2} 
\theta_{l_1+K+1,l_2+K+1} \nonumber\\
& \quad\times \sum_{s_1=0}^{l_1}\frac{d_{s_1+K+1}}{(s_1+K+1)!} \frac{\xi^{s_1}}{(l_1-s_1)!} \nonumber\\
&  \quad\times \sum_{s_2=0}^{l_2}\frac{d_{s_2+K+1}}{(s_2+K+1)!} \frac{\xi^{s_2}}{(l_2-s_2)!} 
 \end{align} 
  As in the estimate of the average value $\langle \Delta P_1\rangle $,  only the lowest-order terms in the sums over $l_{1,2}$ contribute significantly to the result. Thus we can  approximate the  Pochhammer symbols  by using  the expansions   in Eqs. \eqref{Poch_1} and \eqref{Poch_2}. Moreover, for $K\ll \sqrt{N}$, it turns out to be sufficient for the leading-order approximation for $\theta_{n_1,n_2}$ of Eq. \eqref{Aq49}  to keep only the terms with $s=0$ and $s=1$ in the expansion. In this case  one can simplify  the variance $\mathcal{R}$ in Eq. \eqref{Aq50}. Using the following approximations  for the ratios of Pochhammer symbols    (setting $n = \mathrm{max}(n_1,n_2)$)
 \begin{align*}
 &  \frac{M^{(n_1+n_2)}}{M^{(n_1)}M^{(n_2)}} =  1 + \frac{n_1n_2}{M} +  O\left(\frac{n^4}{M^2}\right),\\
 &   \frac{(N)_{(n_1+n_2)}}{M^{(n_1)}M^{(n_2)}} = \left(\frac{N}{M}\right)^{n_1+n_2}\left[1+ O\left(\frac{n^2}{N}\right) \right]
 \end{align*} 
 we obtain for the first two terms ($s=0,1$) in Eq. \eqref{Aq49}
\begin{align*}
&   \sum_{s=0,1}  \frac{(n_1)_s (n_2)_s (N)_{n_1+n_2-s}}{s! M^{(n_1+n_2)}}  \left(2^s - \frac{M^{(n_1+n_2)}}{M^{(n_1)} M^{(n_2)}}\right)\nonumber\\
&  =  \left(\frac{N}{M}\right)^{n_1+n_2}  \left[1+ O\left(\frac{n^2}{N}\right) \right] \nonumber\\
& \times \Biggl\{  - \frac{n_1n_2}{M}    + O\left(\frac{n^4}{M^2}\right) +  \frac{n_1n_2}{N-n_1-n_2+1} \nonumber\\
&  \times \left[ 1 - \frac{n_1n_2}{M} +   O\left(\frac{n^4}{M^2}\right) \right] \Biggr\} \nonumber\\
&  =    \rho^{n_1+n_2}   (1- \rho)   \frac{n_1n_2}{N}\left( 1+ O\left(\frac{n}{N}\right)\right).
\end{align*}
The rest of the expansion in Eq. \eqref{Aq49} is relatively much smaller for $n_{1,2}\le K+1\ll \sqrt{N}$, since  
 \begin{align*}
 & \!\! \!\sum_{s=2}^{\mathrm{min}(n_1,n_2)}   \frac{ (n_1)_s(n_2)_s (N)_{n_1+n_2-s}}{s!M^{(n_1+n_2)}}  \left(2^s - \frac{M^{(n_1+n_2)}}{M^{(n_1)} M^{(n_2)}}\right)\nonumber\\
 &  < \left(\frac{N}{M}\right)^{n_1+n_2}\sum_{s=2}^{\mathrm{min}(n_1,n_2)}   \frac{(2^s-1)(n_1)_s(n_2)_s}{s! N^s}  \nonumber\\
 &  < \rho^{n_1+n_2} \sum_{s=2}^n    \frac{1}{s!} \left( \frac{2n^2}{N} \right)^s \\
 &   =   2 \rho^{n_1+n_2}\frac{n^4}{N^2} \left[ 1+ O\left(\frac{n^2}{N}\right)\right] = O\left( \frac{n^4}{N^2}\right).
\end{align*}
Thus we have (recall that   $n = \mathrm{max}(n_1,n_2)$)
\be
\theta_{n_1,n_2} = \frac{(1-\rho)}{N}  n_1 n_2\rho^{n_1+n_2} \left[1 + O\left(\frac{n^2}{N} \right) \right].
\en{Aq52}

For $K$ sufficiently large, using the approximation $d_m/m!\approx e^{-1}$ for $m\ge K+1$,  we obtain from Eqs. \eqref{Aq50} and  \eqref{Aq52} the leading order term of the variance   as follows 
\begin{align}
\label{Aq53}
&    \mathcal{R}  \approx \frac{ (1-\rho)}{N  }(K+1)^2(\eta \xi \rho)^{2K+2}e^{-2[1+\nu]} \nonumber\\
&  \times   \sum_{l_1,l_2=0}^{N-K-1}  (-\eta\rho)^{l_1+l_2}   \sum_{s_1=0}^{l_1} \frac{\xi^{s_1}}{(l_1-s_1)!}\sum_{s_2=0}^{l_2}\frac{\xi^{s_2}}{(l_2-s_2)!} \nonumber\\
 & =\frac{(1-\rho)(K+1)^2}{N}  \left(\frac{e^{-1-\nu-\eta\rho}(\eta \xi \rho)^{K+1}}{1 + \xi\eta \rho } \right)^2,
\end{align}
where we have taken into account that the four sums  factorize into two double sums for   $(l_i,s_i)$, $i=1,2$, with each factor  being  proportional to that evaluated  in  Eq. \eqref{Aq33}. Eq. \eqref{Aq53}  shows that,  as \mbox{$N\to \infty$} and   $K^2/N \to 0$,  the   average value $\langle \Delta P_1\rangle$,  given by  Eq. \eqref{Aq34},   is   the  asymptotic difference in probability  for almost all interferometers.

Obviously, there are interferometers which cannot satisfy   Eq. \eqref{Aq34}. This  is the  class of almost trivial interferometers $\mathcal{U}= \sqrt{\eta}U$, $UU^\dag=I$, where $U = \mathcal{P}V$ is a product of a permutation interferometer $\mathcal{P}$, exchanging the labels of the input ports, and an almost diagonal unitary interferometer $V$. More precisely, if   $ \mathrm{max}(|V_{k,l}|^2) $ for $k\ne l$ is  much   smaller than the average $\langle |U_{k1}|^2\rangle = 1/M$.  This class of interferometers  corresponds to strongly concentrated   output probability distribution, as $N$ scales up, on the  permutation of input ports by $\mathcal{P}$.

\section{Estimating the probability  $P_1$}
\label{appE}

We consider the case of uniform losses and distinguishability:  $\mathcal{U} = \sqrt{\eta}U$, $U^\dag U = I$, and   $J(\sigma) = \xi^{N-c_1(\sigma)}$.   
In the simplest case of $L=1$ we can  estimate the average value of $P_1$ of Eq. \eqref{B5} for such a noisy boson sampling model. The easiest way to get the necessary  expression for $\langle P_1\rangle $ is by   replacing   \mbox{$K+1\to 0$}  in  Eq. \eqref{Aq31}. We obtain
\begin{align}
\label{E1}
  \langle P_1\rangle & = e^{-\nu} \sum_{n=0}^N \binom{N}{n}  \frac{(-\eta)^n}{M^{(n)}} \sum_{m=0}^{n}\binom{n}{m}d_{m} \xi^{m}\nonumber\\
&= e^{-\nu} \sum_{n=0}^N    \frac{(N)_n}{M^{(n)}} (-\eta)^n\sum_{s=0}^n \frac{\xi^{n-s}(1-\xi)^s}{s!},
\end{align}
where we have used the summation identity (which can be obtained by the method of generating functions, see for instance, Ref. \cite{Stanley})
\[
 \sum_{m=0}^{n}\binom{n}{m}d_{m} \xi^{m} =  n!\sum_{s=0}^n \frac{\xi^{n-s}(1-\xi)^s}{s!}.
\]
Using Eqs. \eqref{Poch_1} and \eqref{Poch_2} we obtain 
\begin{align}
\label{E2}
  \frac{(N)_n}{M^{(n)}} (-\eta)^{n} =   (-\eta\rho)^n   \left[1+ O\left(\frac{n^2}{N}\right) \right].
    \end{align}
Now we can perform an approximation assuming that $\rho\eta\ll1$, such that only  the powers  $n\ll \sqrt{N}$ contribute significantly to the   sum  over $n$ in  Eq. \eqref{E1}, similar as in the computation of the average difference in probability $\langle \Delta P_1\rangle$. 
We get 
\begin{align}
\label{E3}
&   \langle P_1\rangle \approx  e^{-\nu} \sum_{n=0}^\infty (-\xi\eta \rho)^n \sum_{s=0}^n  \frac{(1/\xi - 1)^s}{s!}\nonumber\\
&   = e^{-\nu}\sum_{s=0}^\infty  \frac{(1/\xi - 1)^s}{s!} \sum_{n=s}^\infty (-\xi\eta\rho)^n \nonumber \\
& = \frac{\exp\left(-\nu-\eta\rho[1-\xi]\right)}{1+\xi\eta\rho}.\nonumber\\
\end{align}

Let us also estimate the variance $\mathcal{V}\equiv \langle P_1^2\rangle - \langle P_1\rangle^2$.  We have from Eqs.  \eqref{Aq47} and \eqref{Aq50} by  using   Eqs. \eqref{Aq31}, \eqref{E1}, and \eqref{E3}
\begin{align}
\label{E4}
&  \mathcal{V} \approx \frac{(1-\rho)}{N}\left[ e^{-\nu} \sum_{n=0}^N n (-\rho\eta)^n\sum_{m=0}^n \frac{d_m}{m!}\frac{\xi^m}{(n-m)!} \right]^2\nonumber\\
&  \approx  \frac{(1-\rho)}{N}\left( -\eta \frac{\partial }{\partial \eta} \langle P_1\rangle\right)^2\nonumber\\
&  \approx \frac{(1-\rho)}{N} (\eta\rho)^2\left(1-\xi + \frac{\xi}{1+\rho\eta\xi} \right)^2\langle P_1\rangle^2,
\end{align}
where $\langle P_1\rangle$ is given in Eq. \eqref{E3}.

\section{Numerical simulations  of the lower bound}
\label{appD}

Here the method used to numerically simulate   Eq. \eqref{Eq3} of the main text is described. We assume, as in appendix  \ref{appC},  a uniform overlap $\xi$ over the internal states of bosons, thus $ J(\sigma) = \xi^{N-c_1(\sigma)}$, where $c_1(\sigma)$ is the total number of fixed points of permutation $\sigma$.  Denoting $n = c_1(\sigma)$ we get from    Eqs. \eqref{Aq10} and \eqref{Aq13}
\begin{widetext}
\begin{align}
\label{Aq54}
\Delta P_L& =  e^{-\nu L}\sum_{\sigma\in S_N} \Delta J (\sigma)\prod_{k=1}^N A_{k,\sigma(k)} =  e^{-\nu L}  \sum_{n=0}^{N-K-1}\xi^{N-n} \sum_{\bk}\sum_{\mathrm{fix}(\sigma) = \bk}  \left[\prod_{\alpha=1}^n  A_{k_\alpha,k_\alpha}\right] \prod_{\alpha=n+1}^N A_{k_\alpha,\sigma(k_\alpha)}\nonumber\\
& =  e^{-\nu L}  \sum_{n=0}^{N-K-1} \sum_{\bk} \sum_{\mathrm{fix}(\sigma) = \bk} \left[\prod_{\alpha=1}^n  A^{(\xi)}_{k_\alpha,k_\alpha}\right] \prod_{\alpha=n+1}^N A^{(\xi)}_{k_\alpha,\sigma(k_\alpha)},
\end{align}
\end{widetext}
where    $\bk = (k_1,\ldots,k_n)$,  the sum with $\mathrm{fix}(\sigma)=\bk$ denotes the summation over all permutations with    fixed  points $\bk$ (i.e., over all derangements of the complementary subset $(k_{n+1},\ldots,k_{N})$),     and $A^{(\xi)}_{kl} = \delta_{kl} A_{kk} + \xi (1-\delta_{kl}) A_{kl}$.    Let us define  an additive  function $f(\sigma)$ on  the symmetric group $S_N$ by setting  
\be
f(\sigma) = \prod_{k=1}^N A^{(\xi)}_{k,\sigma(k)}.
\en{Aq55}
Then the  sum in the expression on the r.h.s. of Eq. \eqref{Aq54} can be rewritten as follows
\begin{align}
\label{Aq56}
&  \sum_{n=0}^{N-K-1} \sum_{\bk} \sum_{\mathrm{fix}(\sigma) = \bk}\left[\prod_{\alpha=1}^n A^{(\xi)}_{k_\alpha,k_\alpha}\right] \prod_{\alpha=n+1}^N A^{(\xi)}_{k_\alpha,\sigma(k_\alpha)} \nonumber\\
&  \quad = \sum_{n=0}^{N-K-1}  f(D_{N-n}),
\end{align}
where $D_{m}\in S_N$ is the set of all permutations having exactly $N-m$ fixed points, i.e., the set of all derangements of $m$ elements in the symmetric group $S_N$. Unfortunately,   derangements are very difficult to handle numerically. Therefore, we need to rewrite the sum in Eq. \eqref{Aq56} in terms of the matrix permanents (i.e., using all the  permutations in a symmetric group).  We will use the following summation identity (representing the generalised inclusion-exclusion principle)  valid for any additive function $f$ on the symmetric group $S_N$ \cite{IE} 
\begin{align}
\label{Aq57}
&   f(D_{N-n}) = \sum_{s=n}^N (-1)^{s-n} \binom{s}{n}f_s,\nonumber\\ 
&   f_s \equiv \sum_{k_1\ldots k_s} f(S_{(k_{s+1},\ldots,k_N)}), 
\end{align}
where $S_{(k_{s+1},\ldots,k_N)}$ is the symmetric subgroup $S_{N-s}$ of all permutations of $(k_{s+1},\ldots,k_{N})$. From Eq. \eqref{Aq55}  we  obtain 
\begin{align}
 \label{Aq58}
&   f_s = \sum_{k_1\ldots k_s} \left[\prod_{\alpha=1}^s A^{(\xi)}_{k_\alpha,k_\alpha} \right]\sum_{\sigma\in S_{N-s}}
\prod_{\alpha=s+1}^N A^{(\xi)}_{k_\alpha,\sigma(k_\alpha)} \nonumber\\
&  \!\!\!= \!\!\sum_{k_1\ldots k_s} \!\!\left[\prod_{\alpha=1}^s A^{(\xi)}_{k_\alpha,k_\alpha} \right]\!\mathrm{per}\!\left(\!A^{(\xi)}[k_{s+1}\!\ldots \!k_N|k_{s+1}\!\ldots\! k_N]\right)\!,\nonumber\\
\end{align}
i.e., we need to compute permanents of the positive semi-definite  Hermitian matrices $A^{(\xi)}[k_{s+1}\ldots k_N|k_{s+1}\ldots k_N]$, obtained by  the  rows and columns $k_{s+1},\ldots, k_N$ of $A^{(\xi)}$.
Using Eq. \eqref{Aq58}  can rewrite the sum in Eq. \eqref{Aq56} as follows
\begin{align}
\label{Aq59}
&  \sum_{n=0}^{N-K-1}  f(D_{N-n}) = \sum_{n=0}^{N-K-1} \sum_{s=n}^N (-1)^{s-n} \binom{s}{n} f_s \nonumber\\
&  = \sum_{n=0}^{N-K-1}\left(  \sum_{s=n}^{N-K-1} +  \sum_{s=N-K}^N  \right)(-1)^{s-n} \binom{s}{n} f_s \nonumber\\
&  =  \left(\sum_{s=0}^{N-K-1}\sum_{n=0}^s +  \sum_{s=N-K}^N\sum_{n=0}^{N-K-1}  \right)(-1)^{s-n} \binom{s}{n} f_s\nonumber\\
&  =  f_0 +  (-1)^{N-K-1}\sum_{s=N-K}^{N} (-1)^s  \binom{s-1}{N-K-1}f_s, \nonumber\\
\end{align}
where we have used the  summation identity (easily proven by induction in $m$)
\[
\sum_{n=0}^m(-1)^n\binom{s}{n} = \left\{ \begin{array}{cc} (-1)^m \binom{s-1}{m},  & m<s\\ 
\delta_{s,0}, & m = s. \end{array}  \right.
\]
Therefore, we have derived the following result 
\begin{align}
 \label{Aq60}
\!\!\Delta P_L  \!= \!e^{-\nu L}\left[\!f_0 -\! \!\!\sum_{s=N-K}^{N} \!\!(-1)^{N-K-s} \! \binom{s\!-\!1}{N\!-\!K\!-\!1}f_s\!\right]\!,\nonumber\\
\end{align}
where $f_s$ is given by. Eq. \eqref{Aq58}.  This result allows one to simulate numerically $\Delta P_L$  by computing the matrix permanents.

\section{Comparing the models with distinguishability functions  $J^{(K)}$ and $J_+^{(K)}$}
\label{appF}

\begin{figure}[htb]
\begin{center}
   \includegraphics[width=.25\textwidth]{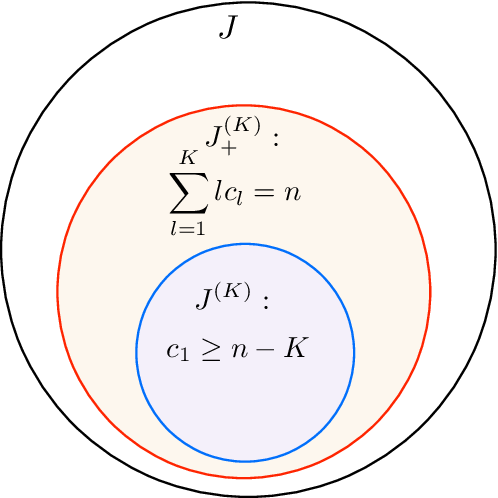} 
       \caption{Schematic view of the relation between the support domains in $S_n$ of the  original distinguishability function  $J(\sigma)$,   the   model  accounting  for all $K$-boson interferences,   $J_+^{(K)}(\sigma)$,  and    the model  of $K$ interfering bosons and $N-K$ classical particles,  $J^{(K)}(\sigma)$.    \label{fig3} }
   \end{center}
\end{figure}

Let us consider how switching from  the approximation model with $J^{(K)}(\sigma)$ of Eq. \eqref{Eq2} to that with $J_+^{(K)}(\sigma)$ of Eq. \eqref{JKnew} would affect the results  in Eqs. \eqref{Eq3}-\eqref{Eq4}. The difference in probability  $\Delta P_L$ in Eq. \eqref{Eq3} depends on the difference of distinguishability functions, moreover, the expression for $\langle \Delta P_1\rangle$ depends only on the sum  $\sum_{\sigma\in S_n}\Delta J(\sigma)$ for $K+1\le n\le N$ (see Eqs. \eqref{Aq15}-\eqref{Aq19}  in appendix \ref{appC}), which for  uniform distinguishability and losses,   read:
\begin{align}
\label{F1}
&  \!\!\sum_{\sigma\in S_n} \left[J (\sigma) -J^{(K)}(\sigma)\right]  \!= \!\sum_{m=K+1}^n \binom{n}{m}\xi^m d_m ,\nonumber\\
& \!\!\! \sum_{\sigma\in S_n} \!\left[J(\sigma) -J_+^{(K)}(\sigma)\! \right]  \!= \!\sum_{m=K+1}^n \!\binom{n}{m}\xi^m(d_m\!-\!d^{(K)}_m),\nonumber\\
\end{align}
where we have introduced the number of derangements $m\equiv n-c_1(\sigma)$, have   taken into account that  derangements of $m$ bosons are weighted by $\xi^m$,  that the  model with $J^{(K)}$ accounts only  for the  derangement  with    $m\le K$, whereas  that with $J_+^{(K)}$     accounts for those   ($d^{(K)}_m$) containing  only the  $l$-cycles with  $l\le K$, see fig. \ref{fig3}.

  To find an expression for $d^{(K)}_m$ one can use the generation function method, which allows to compute any cycle sum $Z_m(t_1,\ldots,t_m)$ over permutation group $S_m$,
\be
Z_m(t_1,\ldots,t_m) = \sum_{\sigma\in S_m} \prod_{l=1}^m t_l^{c_l(\sigma)},
\en{F2}
by using the explicit form of the  generating function \cite{Stanley}
\be
F(X) \equiv \sum_{m\ge 1}Z_m(t_1,\ldots,t_m)X^m = \exp\left( \sum_{l=1}^\infty t_l\frac{X^l}{l}\right),
\en{F3}
with $Z_m(t_1,\ldots,t_m) = (\frac{d}{dX})^m_{X=0}F(X)$. 
Let us first   approximate $d_m$ using this method.  In this case   $t_l=1$ for $l\ge 2$ and $t_1=0$ in Eq. \eqref{F2}.  From  Eq. \eqref{F3}  we obtain 
\begin{align}
\label{F4} 
  d_m &=  \left(\frac{d}{dX}\right)^m_{\! X=0}\exp\left( \sum_{l=2}^\infty \frac{X^l}{l}\right)\nonumber\\
&=   \left(\frac{d}{dX}\right)^m_{\! X=0} \frac{e^{-X}}{1-X}  \nonumber\\
 &= m! \sum_{s=0}^m \frac{1}{s!} \left(\frac{d}{dX}\right)^s_{\! X=0}\exp\left( -X \right)\nonumber\\
&=  m!\left\{ e^{-1} + R_m\right\},
\end{align}
 where we have used the Leibniz formula for the $m$th derivative of a product and    approximated  the Taylor series  of $e^{-X}$  by its value at $X=1$. The remainder $R_m$ 
is   bounded by the $(m+1)$th term of the Taylor series maximized on the interval $[0,1]$:
\be
| R_m |\le  \frac{1}{(m+1)!}. 
 \en{R_m} 
 For   $d^{(K)}_m$  one must set  in $Z_m(t_1,\ldots, t_m)$ of Eq. \eqref{F2}    $t_1= t_{K+i}=0$, for all $i\ge 1$, and  $t_l  =1$ otherwise. Similar manipulations as  in Eq. \eqref{F4}  give 
 \begin{align}
\label{F5}
 d^{(K)}_m  &\!=  \left(\frac{d}{dX}\right)^m_{\! X=0} \exp\left( \sum_{l=2}^K \frac{X^l}{l}\right)\nonumber\\
&  \!=  \left(\frac{d}{dX}\right)^m_{\! X=0} \frac{\exp\left( -X - \sum_{l=K+1}^\infty \frac{X^l}{l}\right)}{1-X}\nonumber\\
&  \! = m! \sum_{s=0}^m \frac{1}{s!} \left(\frac{d}{dX}\right)^s_{\! X=0}\exp\left( -X - \sum_{l=K+1}^\infty \frac{X^l}{l}\right)\nonumber\\
&  \!= m!  \left\{ \exp\left( -1 - \sum_{l=K+1}^{Q_m} \frac{1}{l}\right) +R^{(K)}_{Q_m}\right\},\nonumber\\ 
\end{align}
where for $m\ge K+1$ the optimal $Q_m\ge m$, minimizing the remainder $R^{(K)}_{Q_m}$ is unknown (i.e., the remainder is hard  to estimate). This is the main difficulty of working out  calculations in the model with $J^{(K)}_+$. 

Let us now  comment on what  changes occur  when   our  model with $J^{(K)}$ is substituted by that with  $J_+^{(K)}$.   In the  expression for the average   $|\langle \Delta P_1\rangle|$ in   appendix \ref{appC}    the number of derangements, approximated as  $d_m \approx m! e^{-1}$    (for $m \ge K+1$),   has been used only once,   to   obtain the factor  $\frac{\exp\{-1- \rho \eta \}}{1 + \rho\eta \xi}$ in the expression for    $W_1$  Eq. \eqref{Eq4},   see   Eqs. \eqref{Aq32} and \eqref{Aq33}  of  appendix \ref{appC}.   Hence, the only change  in the result would be in    the   double sum in  Eq. \eqref{Aq32}  with $d_m-d^{(K)}_m$  replacing  $d_m$, whereas  the rest of appendix \ref{appC} does not change. By   Eqs. \eqref{F4} and \eqref{F5} the difference still satisfies a similar scaling \mbox{$d_m - d^{(K)}_m = m! f(m,K)$,} where $f(m,K)$ is a bounded function for all $m\ge K+1$.  Therefore,   instead of $e^{-1}$ the new expression for $W_1$ would get in the case of the model  with $J^{(K)}_+$   a more complicated factor.     
 


\begin{thebibliography}{99}

\bibitem{F} R.  Feynman. Simulating Physics with Computers. \href{https://doi.org/ 10.1007/BF02650179}{\textit{Int. J. Theoret. Phys.} \textbf{21},  467-488 (1982).}

\bibitem{Sh} P. W. Shor. Algorithms for quantum computation: discrete logarithms and factoring. \href{https://doi.org/10.1109/SFCS.1994.365700}{\textit{Proceedings of the 35th Annual Symposium Foundations of Computer Science (IEEE, New York, 1994), p. 124–134.}}


\bibitem{P} J. Preskill. Quantum Computing in the NISQ era and beyond. \href{https://doi.org/10.22331/q-2018-08-06-79}{\textit{Quantum} \textbf{2,} 79 (2018). }



\bibitem{AA} S. Aaronson and A. Arkhipov,  The computational complexity of linear optics.  \href{https://doi.org/10.4086/toc.2013.v009a004}{\textit{Theory of Computing} \textbf{9},  143 (2013).}


\bibitem{QSProp} M. J. Bremner,  A.  Montanaro, and D. J. Shepherd. Achieving quantum supremacy with sparse and noisy commuting quantum computations. 
 \href{https://doi.org/10.22331/q-2017-04-25-8}{\textit{Quantum} \textbf{1}, 8 (2017). }

\bibitem{QSArch}   J.~Bermejo-Vega, D.~Hangleiter,   M. Schwarz, R. Raussendorf,  and J.~Eisert. Architectures for Quantum Simulation Showing a Quantum Speedup. 
 \href{https://doi.org/10.1103/PhysRevX.8.021010}{\textit{ Phys. Rev. X } \textbf{8}, 021010 (2018). }
	
\bibitem{QSChar} S. O. Boixo, S. V. Isakov, V. N. Smelyanskiy, R. Babbush,  N.  Ding, Z. Jiang, M.  J. Bremner, J. M. Martinis, and H. Neven.
Characterizing quantum supremacy in near-term devices. \href{https://doi.org/10.1038/s41567-018-0124-x}{\textit{Nature Physics}, \textbf{14}, 595-600 (2018).}

\bibitem{QSColdAt} X. Gao, S.-T. Wang,  and L.-M. Duan. Quantum Supremacy for Simulating a Translation-Invariant Ising Spin Model. 
\href{https://doi.org/10.1103/PhysRevLett.118.040502}{\textit{ Phys. Rev. Lett. } \textbf{118}, 040502 (2017).} 

\bibitem{GoogleS} F. Arute, K. Arya, R. Babbush, D. Bacon, J. C. Bardin, R. Barends, R. Biswas, S. Boixo, F. G. S. L. Brandao, D. A. Buell, \textit{et al}.
Quantum supremacy using a programmable superconducting processor. \href{https://doi.org/10.1038/s41586-019-1666-5}{\textit{Nature} \textbf{574,}  505-510 (2019).}


\bibitem{HQCF} G. Kalai. The Quantum Computer Puzzle. \href{http://dx.doi.org/10.1090/noti1380}{\textit{ Notices of the AMS}, \textbf{63}, 508-516 (2016). } 


\bibitem{Bbirthday} A. Arkhipov and G. Kuperberg.  The bosonic birthday paradox. 	\href{https://doi.org/10.2140/gtm.2012.18.1}{\textit{Geometry \& Topology Monographs} {\bf 18}, 1-7 (2012).}


\bibitem{C} E. R. Caianiello. On quantum field theory — I: explicit solution of Dyson’s equation in electrodynamics without use of Feynman graphs.
\href{https://doi.org/10.1007/BF02781659}{ \textit{Nuovo Cimento}, \textbf{10}, 1634-1652 (1953)}; 
\textit{Combinatorics and Renormalization in Quantum Field Theory}, Frontiers in Physics, Lecture Note Series (W. A. Benjamin, Reading, MA, 1973).




\bibitem{Scheel} S. Scheel. Permanents in linear optical networks.  \href{https://arxiv.org/abs/quant-ph/0406127}{arXiv:quant-ph/0406127}.  


\bibitem{Valiant} L. G. Valiant.  The complexity of computing the permanent. 
\href{https://doi.org/10.1016/0304-3975(79)90044-6}{\textit{Theoretical Comput. Sci.,} \textbf{8},  189-201 (1979).}

\bibitem{JSV} M. Jerrum,  A.  Sinclair, and E. Vigoda. A polynomial-time approximation algorithm for the permanent of a matrix with nonnegative entries.
\href{https://doi.org/10.1145/1008731.1008738}{\textit{Journal of the ACM} \textbf{51}, 671-697 (2004).}

\bibitem{A1} S. Aaronson. A linear-optical proof that the permanent is $\#$P-hard.  \href{https://doi.org/10.1098/rspa.2011.0232}{\textit{Proc. Roy. Soc. London A}, \textbf{467},   3393–3405 (2011).}

\bibitem{Ryser} H. Ryser, \textit{Combinatorial Mathematics} (Cams Mathematical Monographs, No. 14;
published by The Mathematical Association of America, distributed by John Wiley
and Sons, 1963).




\bibitem{E1} M. A. Broome, A. Fedrizzi, S. Rahimi-Keshari, J. Dove,  S. Aaronson, T. C. Ralph, and  A. G. White. 
Photonic Boson Sampling in a Tunable Circuit. 
\href{https://doi.org/10.1126/science.1231440}{\textit{Science} \textbf{339},   794-798 (2013).}

\bibitem{E2} J. B. Spring,  B. J. Metcalf, P. C. Humphreys, W.~S.~Kolthammer, X.-M. Jin, M. Barbieri, A.~Datta, N.~Thomas-Peter, N. K. Langford, D. Kundys, J.~C.~Gates,  B. J. Smith, P. G. R. Smith, and  I. A. Walmsley. Boson Sampling on a Photonic Chip. \href{https://doi.org/10.1126/science.1231692}{\textit{Science}, \textbf{339},  798-801 (2013).}

\bibitem{E3}    M. Tillmann, B. Daki\'c, R. Heilmann, S.~Nolte, A. Szameit, and P. Walther. Experimental boson sampling.  \href{https://doi.org/10.1038/nphoton.2013.102}{\textit{Nature Photonics},  \textbf{7}, 540-544 (2013).}

\bibitem{E4} A. Crespi, R. Osellame, R. Ramponi, D.~J.~Brod, E.~F.~Galv\~ao, N. Spagnolo, C. Vitelli, E. Maiorino, P.~Mataloni, and F.~Sciarrino. Integrated multimode interferometers with arbitrary designs for photonic boson sampling.    \href{https://doi.org/10.1038/nphoton.2013.112}{\textit{Nature Photonics}, \textbf{7}, 545-549 (2013).}

\bibitem{EVCC} J. Carolan,  J. D. A. Meinecke, P. J. Shadbolt, N.  J. Russell, N. Ismail, K. W\"orhoff, T. Rudolph, M. G. Thompson, J. L. O'Brien, J. C. F. Matthews, and  A. Laing. On the experimental verification of quantum complexity in linear optics. \href{https://doi.org/10.1038/nphoton.2014.152}{\textit{Nature Photonics}, \textbf{8},  621-626 (2014). }

\bibitem{GBS}  A. P.  Lund, A. Laing, S. Rahimi-Keshari, T. Rudolph, J. L.  O'Brien, and T. C. Ralph. Boson Sampling from a Gaussian State. 
\href{https://doi.org/10.1103/PhysRevLett.113.100502}{\textit{Phys. Rev. Lett.} \textbf{113}, 100502 (2014). }
 


\bibitem{E5}  M. Bentivegna, N. Spagnolo,  C. Vitelli, F. Flamini, N. Viggianiello, L. Latmiral, P. Mataloni, D. J. Brod, E. F. Galvão, A.  Crespi, R. Ramponi, R. Osellame,  and F. Sciarrino. Experimental scattershot boson sampling. \href{https://doi.org/10.1126/sciadv.1400255}{\textit{Science  Advances} \textbf{1}, e1400255 (2015). }

\bibitem{GSNEW} H.-S. Zhong, L.-C. Peng, Y. Li, Y. Hu, W.  Li, J. Qin, D. Wu, W. Zhang, H. Li, L. Zhang, Z. Wang \textit{et al}.  Experimental Gaussian Boson sampling.
 \href{https://doi.org/10.1016/j.scib.2019.04.007}{\textit{Science Bulletin}, \textbf{64}, 511-515 (2019).}
  
\bibitem{TbinBS} K. R. Motes, A. Gilchrist, J. P. Dowling, and P. P. Rohde. Scalable Boson Sampling with Time-Bin Encoding Using a Loop-Based Architecture. 
\href{https://doi.org/10.1103/PhysRevLett.113.120501}{\textit{Phys. Rev. Lett.} \textbf{113}, 120501 (2014). }

\bibitem{TbinBSExp}  Y. He, X. Ding, Z. E. Su, H. L. Huang, J. Qin, C. Wang, S.
Unsleber, C. Chen, H. Wang, Y. M. He, \textit{et al}. Time-Bin-Encoded Boson Sampling with a Single-Photon Device. \href{https://doi.org/10.1103/PhysRevLett.118.190501}{\textit{Phys. Rev. Lett.} \textbf{118}, 190501  (2017). }



\bibitem{pure1phBS} J. C. Loredo, M. A. Broome, P. Hilaire, O. Gazzano, I. Sagnes, A. Lemaitre, M. P. Almeida, P. Senellart, and A. G. White. Boson Sampling with Single-Photon Fock States from a Bright Solid-State Source. \href{https://doi.org/10.1103/PhysRevLett.118.130503}{\textit{Phys. Rev. Lett.} \textbf{118}, 130503 (2017).}

\bibitem{HEBS} H. Wang, Y. He, Y.-H. Li, Z.-E. Su, B. Li, H.-L. Huang, X. Ding, M.-C. Chen, C. Liu, J. Qin \textit{et al}. High-efficiency multiphoton boson sampling.
\href{https://doi.org/10.1038/nphoton.2017.63}{\textit{Nature Photonics}  \textbf{11},   361-365  (2017). }

\bibitem{LossBS} H. Wang, W. Li, X. Jiang, Y. M. He, Y. H. Li, X. Ding, M. C. Chen, J. Qin, C. Z. Peng, C. Schneider  \textit{et al}. Toward Scalable Boson Sampling with Photon Loss.
\href{https://doi.org/10.1103/PhysRevLett.120.230502}{\textit{Phys. Rev. Lett.} \textbf{120}, 230502 (2018). }

\bibitem{12phBS} H.-S. Zhong,  Y. Li, W. Li, L.-C. Peng, Z.-E. Su, Y. Hu, Y.-M. He, X. Ding, W. Zhang, H.  Li \textit{et al}. 12-Photon Entanglement and Scalable Scattershot Boson Sampling with Optimal Entangled-Photon Pairs from Parametric Down-Conversion. \href{https://doi.org/10.1103/PhysRevLett.121.250505}{\textit{Phys. Rev. Lett.} \textbf{121}, 250505 (2018). }
  


\bibitem{20ph60mod} H.  Wang, J.  Qin, X. Ding, M.-C. Chen, S. Chen, X. You, Y.-M. He, X. Jiang, L. You, Z. Wang, C. Schneider, J. J. Renema, S.~Höfling, C.-Y. Lu, and J.-W. Pan.
 Boson Sampling with 20 Input Photons and a 60-Mode Interferometer in a $10^{14}$-Dimensional Hilbert Space. 
 \href{https://doi.org/10.1103/PhysRevLett.123.250503}{\textit{Phys. Rev. Lett.} \textbf{123},  250503 (2019).} 


\bibitem{BSions} C. Shen, Z. Zhang, and L.-M. Duan. Scalable Implementation of Boson Sampling with Trapped Ions. \href{https://doi.org/10.1103/PhysRevLett.112.050504}{\textit{Phys. Rev. Lett.} \textbf{112}, 050504 (2014).}


\bibitem{BSsuperc}  B. Peropadre, G. G. Guerreschi, J. Huh, and A.  Aspuru-Guzik. Proposal for Microwave Boson Sampling. 
\href{https://doi.org/10.1103/PhysRevLett.117.140505}{\textit{Phys. Rev. Lett.} \textbf{117}, 140505 (2016).} 

\bibitem{Goldstein} S. Goldstein, S. Korenblit, Y. Bendor, H. You, M. R. Geller, and N. Katz. Decoherence and interferometric sensitivity of boson sampling in superconducting resonator networks. \href{https://doi.org/10.1103/PhysRevB.95.020502}{\textit{Phys. Rev. B}   \textbf{95}, 020502(R) (2017).}

\bibitem{BSoptlatt}   A. Deshpande, B. Fefferman, M.~C. Tran, M.~Foss-Feig  and A. V. Gorshkov. Dynamical Phase Transitions in Sampling Complexity.
\href{https://doi.org/10.1103/PhysRevLett.121.030501}{\textit{Phys. Rev. Lett. } \textbf{121}, 030501 (2018).}

\bibitem{BScas}  B. Peropadre, J. Huk and  C.  Sab\'in. Dynamical Casimir Effect for Gaussian Boson Sampling. 
\href{https://doi.org/10.1038/s41598-018-22086-2}{\textit{Scientific Reports} \textbf{8},  3751 (2018).}

\bibitem{QSBS} A. Neville, C. Sparrow, R. Clifford, E. Johnston, P. M. Birchall, A. Montanaro, and A.~Laing. Classical boson sampling algorithms with superior performance to near-term experiments. \href{https://doi.org/10.1038/nphys4270}{\textit{Nature Physics} \textbf{13}, 1153-1157 (2017).}
 
\bibitem{Cliffords} P. Clifford, and R. Clifford. The Classical Complexity of Boson Sampling. 
\href{https://doi.org/10.1137/1.9781611975031.10}{\textit{Proceedings of the 2018 Annual ACM-SIAM Symposium on Discrete Algorithms} pp. 146–55.} 


\bibitem{KK}  G. Kalai and G. Kindler. Gaussian Noise Sensitivity and BosonSampling. \href{https://arxiv.org/abs/1409.3093}{arXiv:1409.3093 [quant-ph].} 

\bibitem{LP} A. Leverrier and R. Garc{\'i}a-Patr{\'o}n. Analysis of circuit imperfections in BosonSampling. 
\href{https://dl.acm.org/doi/abs/10.5555/2871401.2871409}{\textit{Quant. Inf. \& Computation} \textbf{15}, 489-512 (2015). }

\bibitem{VS14}      V. S. Shchesnovich.  Sufficient condition for the mode mismatch of single photons for scalability of the boson-sampling computer.
\href{https://doi.org/10.1103/PhysRevA.89.022333}{\textit{Phys. Rev.  A}  \textbf{89},   022333  (2014).}

 
\bibitem{Arkh}  A. Arkhipov. BosonSampling is robust against small errors in the network matrix. 
\href{https://doi.org/10.1103/PhysRevA.92.062326}{\textit{Phys. Rev.  A}  \textbf{92}, 062326 (2015).}

\bibitem{Brod} S. Aaronson and D. J. Brod. BosonSampling with lost photons. 
\href{https://doi.org/10.1103/PhysRevA.93.012335}{\textit{Phys. Rev.  A}  \textbf{93}, 012335 (2016).}

\bibitem{Latm} L.  Latmiral,   N. Spagnolo and F. Sciarrino. Towards quantum supremacy with lossy scattershot boson sampling.
\href{https://doi.org/10.1088/1367-2630/18/11/113008}{\textit{New J. Phys.} \textbf{18},  113008 (2016).}

\bibitem{RR} P. P. Rohde and T. C. Ralph. Error tolerance of the boson-sampling model for linear optics quantum computing.
\href{https://doi.org/10.1103/PhysRevA.85.022332}{\textit{Phys. Rev. A} \textbf{85}, 022332 (2012).}

\bibitem{K1} S. Rahimi-Keshari,  T.  C. Ralph, and C.  M. Caves. Sufficient Conditions for Efficient Classical Simulation of Quantum Optics.
\href{https://doi.org/10.1103/PhysRevX.6.021039}{\textit{Phys. Rev. X} \textbf{6}, 021039 (2016).} 

\bibitem{R1}   J. J. Renema, A. Menssen, W. R. Clements, G. Triginer, W. S. Kolthammer, and I.~A.~Walmsley. Efficient Classical Algorithm for Boson Sampling with Partially Distinguishable Photons. \href{https://doi.org/10.1103/PhysRevLett.120.220502}{\textit{Phys. Rev. Lett.} \textbf{120}, 220502  (2018).}

\bibitem{OB} M. Oszmaniec and D. J. Brod. Classical simulation of photonic linear optics with lost particles. 
\href{https://doi.org/10.1088/1367-2630/aadfa8}{\textit{New J. Phys.} \textbf{20}, 092002 (2018).}

\bibitem{PRS} R. Garc\'ia-Patr\'on, J. J. Renema, and V.~S.~Shchesnovich. Simulating boson sampling in lossy architectures.
\href{https://doi.org/10.22331/q-2019-08-05-169}{\textit{Quantum} \textbf{3}, 169 (2019).}

\bibitem{NonUnifLoss} 	  D. J. Brod and M. Oszmaniec. Classical simulation of linear optics subject to nonuniform losses.
\href{https://doi.org/10.22331/q-2020-05-14-267}{\textit{Quantum} \textbf{4},  267 (2020).}

\bibitem{RSP}  J. J. Renema, V. S. Shchesnovich, and R. Garc\'ia-Patr\'on. Classical simulability of noisy boson sampling.
\href{https://arxiv.org/abs/1809.01953}{arXiv:1809.01953 [quant-ph].} 
 
\bibitem{VS2019} V. S. Shchesnovich. Noise in boson sampling and the threshold of efficient classical simulatability.
\href{https://doi.org/10.1103/PhysRevA.100.012340}{\textit{Phys. Rev. A} \textbf{100}, 012340  (2019).}


\bibitem{BSNotUn}   S. Aaronson and A. Arkhipov. Bosonsampling is far from uniform.
\href{https://dl.acm.org/doi/10.5555/2685179.2685186}{\textit{Quant. Inform.  \& Computation} \textbf{14}, 1383  (2014). } 

\bibitem{BSUn} C. Gogolin, M. Kliesch, L. Aolita, and J. Eisert. Boson-Sampling in the light of sample complexity.
\href{https://arxiv.org/abs/1306.3995}{arXiv:1306.3995 [quant-ph].}


\bibitem{BB} V. S. Shchesnovich. Universality of Generalized Bunching and Efficient Assessment of Boson Sampling.
\href{https://doi.org/10.1103/PhysRevLett.116.123601}{\textit{Phys. Rev. Lett.} \textbf{116}, 123601  (2016).}


\bibitem{StatBench} M.  Walschaers, J. Kuipers, J.-D. Urbina, K. Mayer, M. C. Tichy, K. Richter, and A. Buchleitner.
Statistical benchmark for BosonSampling. \href{https://doi.org/10.1088/1367-2630/18/3/032001}{\textit{New J. Phys.} \textbf{18},  032001 (2016).}

\bibitem{ExpStatSign} T. Giordani, F. Flamini, M. Pompili, N. Viggianiello, N. Spagnolo,
A. Crespi, R. Osellame, N.  Wiebe, M. Walschaers, A. Buchleitner,  and F. Sciarrino. Experimental statistical signature of many-body quantum interference.
\href{https://doi.org/10.1038/s41566-018-0097-4}{\textit{Nature Photonics} \textbf{12,} 173-178 (2018).} 



\bibitem{WDBubles} S. T. Wang and L.-M. Duan. Certification of Boson Sampling Devices with Coarse-Grained Measurements.
\href{https://arxiv.org/abs/1601.02627}{arXiv:1601.02627 [quant-ph].}


\bibitem{PatternRec} I. Agresti, N.  Viggianiello,  F.  Flamini,  N.  Spagnolo,  A.  Crespi, R. Osellame,  N. Wiebe, and F. Sciarrino.
Pattern Recognition Techniques for Boson Sampling Validation. \href{https://doi.org/10.1103/PhysRevX.9.011013}{\textit{Phys. Rev. X} \textbf{9,} 011013 (2019).}


\bibitem{MyEst} V. S. Shchesnovich. On the classical complexity of sampling from quantum interference of indistinguishable bosons.
\href{https://doi.org/10.1142/S0219749920500446}{\textit{Int. J. of Quantum Inform.} \textbf{18}, 2050044 (2020).} 

\bibitem{Barv} A. I. Barvinok. Two Algorithmic Results for the Traveling Salesman Problem.
\href{https://doi.org/10.1287/moor.21.1.65}{\textit{Math. of Oper. Research}, \textbf{21} 65-84 (1996);} see theorem (3.3).

\bibitem{CompHard} V. S. Shchesnovich. Asymptotic evaluation of bosonic probability amplitudes in linear unitary networks in the case of large number of bosons. 
\href{https://doi.org/10.1142/S0219749913500457}{\textit{Int. J. Quantum Inform.} \textbf{11},  1350045 (2013);} see appendix D. 

\bibitem{MRRT} A. E. Moylett,  R. Garc\'ia-Patr\'on, J. J. Renema, and P. S. Turner. Classically simulating near-term partially-distinguishable and lossy boson sampling.
\href{https://doi.org/10.1088/2058-9565/ab5555}{\textit{Quantum Sci. Technol.}  \textbf{5},  015001  (2020).}

\bibitem{S2020}   A. L. Migdall, D. Branning, and S. Castelletto. Tailoring single-photon and multiphoton probabilities of a single-photon on-demand source.
\href{https://doi.org/10.1103/PhysRevA.66.053805}{\textit{Phys. Rev. A}  \textbf{66}, 053805. (2002).}

\bibitem{LossNet} S. M. Barnett, C. R. Gilson, B. Huttner, and N. Imoto. Field Commutation Relations in Optical Cavities.
\href{https://doi.org/10.1103/PhysRevLett.77.1739}{\textit{Phys. Rev. Lett.} \textbf{77}, 1739  (1996).} 

\bibitem{HOM} C. K. Hong, Z. Y. Ou, and L. Mandel. Measurement of subpicosecond time intervals between two photons by interference.
\href{https://doi.org/10.1103/PhysRevLett.59.2044}{\textit{Phys. Rev. Lett.} \textbf{59},   2044 (1987).}

 \bibitem{PartDist} V. S. Shchesnovich. Partial indistinguishability theory for multiphoton experiments in multiport devices. 
 \href{https://doi.org/10.1103/PhysRevA.91.013844}{\textit{Phys. Rev. A}  \textbf{91}, 013844  (2015).}



\bibitem{Stanley}   R. P. Stanley, \textit{Enumerative Combinatorics}, 2nd ed., Vol. 1 (Cambridge University Press, 2011). 

\bibitem{Ninter} V. S. Shchesnovich and  M. E. O. Bezerra. Collective phases of identical particles interfering on linear multiports.
\href{https://doi.org/10.1103/PhysRevA.98.033805}{\textit{Phys. Rev. A} \textbf{98}, 033805 (2018).}

\bibitem{VS2020}  V. S. Shchesnovich and M. E. O. Bezerra. Distinguishability theory for time-resolved photodetection and boson sampling.
\href{https://doi.org/10.1103/PhysRevA.101.053853}{\textit{Phys. Rev. A} \textbf{101},  053853 (2020).} 

\bibitem{HaarR} Z. Puchala  and  J. A.  Miszczak. Symbolic integration with respect to the Haar measure
on the unitary groups. \href{https://doi.org/10.1515/bpasts-2017-0003}{\textit{Bull. Polish Acad. Sci.: Techn. Sci.} \textbf{65}, 21-27 (2017).}


\bibitem{VS2017}  V. S. Shchesnovich. Asymptotic Gaussian law for noninteracting indistinguishable particles in random networks. 
\href{https://doi.org/10.1038/s41598-017-00044-8}{\textit{Scientific Reports}  \textbf{7},  31  (2017).}


\bibitem{QuantOptPerHerm} S. Rahimi-Keshari, A. P. Lund, and T. C. Ralph. What Can Quantum Optics Say about Computational Complexity Theory?
\href{https://doi.org/10.1103/PhysRevLett.114.060501}{\textit{Phys. Rev. Lett.} \textbf{114}, 060501 (2015).} 

\bibitem{PermThermStates} L. Chakhmakhchyan, N. J. Cerf, and R.~Garc\'ia-Patr\'on.
Quantum-inspired algorithm for estimating the permanent of positive semidefinite matrices. 
\href{https://doi.org/10.1103/PhysRevA.96.022329}{\textit{Phys. Rev.  A} \textbf{96}, 022329 (2017).} 


\bibitem{EstTheory} A.  Agresti and  B. A. Coull. Approximate is Better than “Exact” for Interval Estimation of Binomial Proportions.
\href{https://doi.org/10.1080/00031305.1998.10480550}{\textit{The American Statistician} \textbf{52}, 119-126 (1998).}


\bibitem{Bogoliubov}  N. N. Bogolyubov and N. N. Bogolyubov (Jr.), \textit{Introduction to Quantum Statistical Mechanics}  (Nauka, Moscow (1984)).


 \bibitem{BEC1} M. N. Anderson, J. R. Ensher, M. R. Mathews, C. E. Wieman and E. A. Cornell. Observation of Bose-Einstein Condensation in a Dilute Atomic Vapor.
 \href{https://doi.org/10.1126/science.269.5221.198}{\textit{Science} \textbf{269}, 198-201 (1995).}

\bibitem{BEC2}   K. B. Davis, M.-O. Mewes, M. R. Andrews, N. J. van Druten, D. S. Durfee, D. M. Kurn, and W.~Ketterle.  Bose-Einstein Condensation in a Gas of Sodium Atoms. \href{https://doi.org/10.1103/PhysRevLett.75.3969}{\textit{Phys. Rev. Lett.} \textbf{75}, 3969 (1995).}

\bibitem{Pitaevskii}  L. P. Pitaevskii. Vortex lines in an imperfect Bose gas. \href{http://jetp.ac.ru/cgi-bin/dn/e_013_02_0451.pdf}{\textit{Soviet Phys. JETP} \textbf{13}, 451-454 (1961).}

\bibitem{Gross}  E. P. Gross. Structure of a quantized vortex in boson systems. \href{https://doi.org/10.1007/BF02731494}{\textit{Il Nuovo Cimento} \textbf{20}, 454-477 (1961).}

\bibitem{IE}  L. Tak\'acs.  On the Method of Inclusion and Exclusion. \href{https://doi.org/10.2307/2282913}{\textit{J. of   Amer. Stat.  Assoc.} \textbf{62},  102-113 (1967).}

\end{thebibliography}
\end{document}